%% file: ms.tex
\documentclass[iop,apj]{emulateapj}

\usepackage{natbib}
\usepackage{xspace}
\usepackage{amsmath}
\usepackage{url}
\usepackage{apjfonts}
\usepackage{aas_macros}
\usepackage{hyperref}
\hypersetup{colorlinks,citecolor=blue,urlcolor=blue,linkcolor=blue}

\newcommand{\myemail}{\href{mailto:sugayu@icrr.u-tokyo.ac.jp}{sugayu@icrr.u-tokyo.ac.jp}}

\newcommand{\fibsfr}{{$\mathrm{SFR_{fiber}}$}}

\newcommand{\zz}{$z$0\xspace}
\newcommand{\zo}{$z$1\xspace}
\newcommand{\zt}{$z$2\xspace}
\newcommand{\rest}{_{\mathrm{rest}}}
\newcommand{\sys}{_{\mathrm{sys}}}
\newcommand{\osys}[1]{_{\mathrm{#1}, \mathrm{sys}}}
\newcommand{\out}{_{\mathrm{out}}}
\newcommand{\oout}[1]{_{\mathrm{#1}, \mathrm{out}}}
\newcommand{\obs}{_{\mathrm{obs}}}
\newcommand{\cont}{_{\mathrm{cont}}}
\newcommand{\intr}{_{\mathrm{int}}}
\newcommand{\emi}{_{\mathrm{emi}}}

\newcommand{\maxi}{_{\mathrm{max}}}
\newcommand{\SFR}{{\mathrm{SFR}}}
\newcommand{\cir}{_{\mathrm{cir}}}

\newcommand{\lamlam}{$\lambda\lambda$}
\newcommand{\NaID}{\mbox{Na\,{\sc i}\,D}\xspace}
\newcommand{\HeI}{\mbox{He\,{\sc i}}\xspace}
\newcommand{\MgI}{\mbox{Mg\,{\sc i}}\xspace}

\newcommand{\MgII}{\mbox{Mg\,{\sc ii}}\xspace}

\newcommand{\SiII}{\mbox{Si\,{\sc ii}}\xspace}
\newcommand{\SiIIe}{\mbox{Si\,{\sc ii}*}\xspace}

\newcommand{\CII}{\mbox{C\,{\sc ii}}\xspace}
\newcommand{\CIV}{\mbox{C\,{\sc iv}}\xspace}
\newcommand{\OIII}{\mbox{O\,{\sc iii}}}
\newcommand{\NII}{\mbox{N\,{\sc ii}}}
\newcommand{\Xn}{\mbox{$\mathrm{X}^n$}\xspace}
\newcommand{\FeII}{\mbox{Fe\,{\sc ii}}\xspace}
\newcommand{\sub}[1]{_\mathrm{#1}}
\newcommand{\msub}[1]{$_\mathrm{#1}$}
\newcommand{\eX}{\mathrm{X}}
\newcommand{\eH}{\mathrm{H}}

\newcommand{\cgas}{M_\mathrm{gas}^\mathrm{cool}}

\newcommand{\voZz}{145}
\newcommand{\voZzerr}{12}
\newcommand{\voZzerrfix}{2}
\newcommand{\aoZz}{0.03}
\newcommand{\aoZzerr}{0.03}
\newcommand{\vmZz}{174}
\newcommand{\vmZzerr}{9}
\newcommand{\vmZzerrfix}{2}
\newcommand{\amZz}{0.25}
\newcommand{\amZzerr}{0.04}
\newcommand{\vomgi}{163}
\newcommand{\vomgierr}{18}
\newcommand{\vmmgi}{227}
\newcommand{\vmmgierr}{15}

\newcommand{\vomgii}{251}
\newcommand{\vomgiierr}{3}
\newcommand{\vmmgii}{241}
\newcommand{\vmmgiierr}{4}

\newcommand{\vz}{0.59}
\newcommand{\vzerr}{0.03}

\newcommand{\eEtaZz}{2.9}
\newcommand{\eEtaZzerr}{16.3}
\newcommand{\eEtaZzerrf}{0.5}
\newcommand{\eAZz}{-0.2}
\newcommand{\eAZzerr}{1.1}
\newcommand{\eEtaZo}{6.3}
\newcommand{\eEtaZoerr}{4.5}
\newcommand{\eEtaZt}{11.3}
\newcommand{\eEtaZterr}{3.6}
\newcommand{\eZ}{1.2}
\newcommand{\eZerr}{0.3}

\shorttitle{Evolution of Outflow at $z\sim0$--2}
\shortauthors{Sugahara et al.}
\slugcomment{ApJ in press}

\defcitealias{Bruzual:2003}{BC03}
\defcitealias{Maraston:2011a}{MS11}

\begin{document}

\title{Evolution of Galactic Outflows at $\lowercase{z}\sim0$--2 Revealed with SDSS, DEEP2, and Keck spectra}

\author{Yuma Sugahara\altaffilmark{1,2}, 
Masami Ouchi\altaffilmark{1,3}, 
Lihwai Lin\altaffilmark{4},
Crystal L. Martin\altaffilmark{5}, 
Yoshiaki Ono\altaffilmark{1}, 
Yuichi Harikane\altaffilmark{1,2}, \\
Takatoshi Shibuya\altaffilmark{1}, 
and
Renbin Yan\altaffilmark{6}}

\affil{$^1$Institute for Cosmic Ray Research, The University of Tokyo, 5-1-5 Kashiwanoha, Kashiwa, Chiba 277-8582, Japan; \myemail}
\affil{$^2$Department of Physics, Graduate School of Science, The University of Tokyo, 7-3-1 Hongo, Bunkyo, Tokyo, 113-0033, Japan}
\affil{$^3$ Kavli Institute for the Physics and Mathematics of the Universe (WPI), The University of Tokyo, 5-1-5 Kashiwanoha, Kashiwa, Chiba 277-8583, Japan}
\affil{$^4$ Institute of Astronomy \& Astrophysics, Academia Sinica, Taipei 106, Taiwan (R.O.C.)}
\affil{$^5$ Department of Physics, University of California, Santa Barbara, CA, 93106, USA}
\affil{$^6$ Department of Physics and Astronomy, University of Kentucky, 505 Rose St., Lexington, KY 40506-0057, USA}

\begin{abstract}
We conduct a systematic study of galactic outflows in star-forming galaxies
at $z\sim0$--2 based on the absorption lines of optical spectra taken
from SDSS DR7, DEEP2 DR4, and Keck Erb et al.
We carefully make stacked spectra of homogeneous galaxy samples with
similar stellar mass distributions at $z\sim0$--2, and perform the multi-component fitting of 
model absorption lines and stellar continua to the stacked spectra. 
We obtain the maximum ($v\maxi$) and central ($v\out$) outflow velocities, and 
estimate the mass loading factors ($\eta$), a ratio of the mass outflow rate to 
the star formation rate (SFR). 
Investigating the redshift evolution of the outflow velocities
measured with the absorption lines whose depths and ionization energies
are similar (\NaID and \MgI at $z\sim0$--$1$; \MgII and \CII at $z\sim1$--$2$),
we identify, for the first time, that the average value of $v\maxi$ ($v\out$)
significantly increases by 0.05--0.3 dex 
from $z\sim0$ to $2$ at a given SFR.
Moreover, we find that the value of $\eta$ increases 
from $z\sim0$ to 2 by $\eta\propto(1+z)^{\eZ\pm\eZerr}$ 
at a given halo circular velocity $v\cir$, albeit with
a potential systematics caused by model parameter choices.
The redshift evolution of $v\maxi$ ($v\out$) and $\eta$
is consistent with the galaxy-size evolution and
the local velocity-SFR surface density relation, and
explained by high-gas fractions in high-redshift massive galaxies,
which is supported by recent radio observations.
We obtain a scaling relation of 
$\eta\propto v\cir^a$ for $a=\eAZz\pm\eAZzerr$ 
in our $z\sim0$ galaxies that agrees
with the momentum-driven outflow model ($a=-1$)
within the uncertainty.
\end{abstract}

\keywords{galaxies: formation --- 
galaxies: evolution --- 
galaxies: ISM --- 
galaxies: kinematics and dynamics}

\section{INTRODUCTION}
\label{sec:1}
The large scale structure of the universe is well explained 
by gravitational interactions 
based on the Lambda cold dark matter ($\Lambda$CDM) model 
\citep[e.g.,][]{Springel:2005a}.
However, we need additional models to explain 
small scale structures like galaxies, 
because these small scales are significantly affected by the baryon physics, 
such as gas cooling, radiation heating, star formation, and supernovae (SNe).
In fact, 
there is a notable discrepancy 
between the shapes of the halo mass function predicted by numerical simulations of 
$\Lambda$CDM model 
and the galaxy stellar mass function confirmed by observations 
\citep[e.g.,][]{Somerville:2015}.
Theoretical studies suggest that 
this problem of the discrepancy is resolved 
by the feedback processes that regulate the star formation.
Understanding the feedback processes is 
key to galaxy formation and evolution.

Galactic outflows driven by star-forming activities 
are one of the plausible sources of the feedback processes.
The cold gas is accelerated by the starburst or the SNe,  
and carried even to the outside of the halos.
The lack of the cold gas thus quenches 
the star formation in the galaxies.
Although the outflows make an impact on the star formation activity, 
their physical mechanisms are still poorly known.
Theoretical studies propose 
some physical processes to launch the galactic outflows 
such as the thermal pressure of the core-collapse SNe 
\citep[e.g.,][]{Larson:1974b, Chevalier:1985, Springel:2003a}, 
the radiation pressure of the starburst 
\citep[e.g.,][]{Murray:2005, Oppenheimer:2006a}, 
and the cosmic ray pressure
\citep[e.g.,][]{Ipavich:1975a, Breitschwerdt:1991a, Wadepuhl:2011a}.

Since the outflow scale is very small in cosmological simulations, 
major numerical simulations carry out the outflows as subgrid physics.
\citep[see][for a recent review]{Somerville:2015}.
On the other hand, some recent simulations employ 
the explicit thermal feedback generated by SNe 
with no outflows in subgrid physics, and 
describe the relation between the outflow and galaxy properties 
\citep{Schaye:2015a, Muratov:2015, Barai:2015}.
\citet{Muratov:2015} and \citet{Barai:2015} predict that 
outflow properties evolve towards high redshift 
based on their numerical models.
\citet{Mitra:2015} support the evolution of the outflow properties 
with their analytic model including empirical relations.

Optical and ultra-violet (UV) observations investigate 
the galactic outflows with emission lines 
\citep{Lehnert:1996a, Lehnert:1996b, Heckman:1990, 
  Martin:1998a}, 
and absorption lines 
\citep{Heckman:2000, Schwartz:2004, 
  Martin:2005, Rupke:2005a, Rupke:2005b, Tremonti:2007, Martin:2009, 
  Grimes:2009, Alexandroff:2015}
that are found in the spectra of outflow galaxies 
(``down-the-barrel'' technique).
Particularly, 
the absorption lines are used to probe 
outflow velocities of star-forming galaxies 
at $z\sim0$ 
\citep{Chen:2010, Chisholm:2015a, Chisholm:2016a}, 
at $z\sim1$ 
\citep{Sato:2009, Weiner:2009, Martin:2012, Kornei:2012, Rubin:2014, Du:2016a}, 
and at $z>2$ 
\citep{Shapley:2003, Steidel:2010, Erb:2012, Law:2012b, Jones:2013, Shibuya:2014}.
Moreover, 
absorption lines of background quasars are used to study the galactic outflows of 
foreground galaxies on the sight lines of the background quasars
\citep{Bouche:2012, Kacprzak:2014a, Muzahid:2015b, 
  Schroetter:2015, Schroetter:2016a}.

Despite many observations in a wide redshift range, 
it is challenging to study evolution of outflow velocities 
because of possible systematic errors included in different redshift samples.
The literature uses various procedures to measure outflow velocities, such as 
non-parametric \citep{Weiner:2009, Heckman:2015}, 
one-component \citep{Steidel:2010, Kornei:2012, Shibuya:2014, Du:2016a}, 
and two-component methods \citep{Martin:2005, Chen:2010, Martin:2012, Rubin:2014}.
Moreover, 
we should compare the samples of galaxies 
in the same stellar mass and star formation rate (SFR) ranges
because the outflow properties depend on the stellar mass and SFR 
\citep[i.e.,][]{Weiner:2009, Erb:2012, Heckman:2015}.
Although \citet{Du:2016a} compare 
outflow velocities of star-forming galaxies at $z\sim1$ with those at $z\sim3$ 
that are derived with the same procedure, 
\citet{Du:2016a} cannot make similar galaxy samples at $z\sim1$ and $z\sim3$ 
due to the lack of stellar mass measurements.

In this paper, 
we investigate the redshift evolution of galactic outflows 
found in the star-forming galaxies. 
We use spectra of galaxies at $z\sim0$, 1, and 2 
drawn from 
the Sloan Digital Sky Survey 
\citep[SDSS;][]{York:2000}, 
the Deep Evolutionary Exploratory Probe 2 (DEEP2) Galaxy Redshift Survey 
\citep{Davis:2003, Davis:2007, Newman:2013}, and 
\citet{Erb:2006c, Erb:2006b}, respectively.
The combination of these data sets enables us 
to study the redshift evolution of outflow velocities 
based on the samples of star-forming galaxies 
with the same  stellar mass and SFR ranges.
Section \ref{sec:2} presents three samples of star-forming galaxies 
and our methods for spectrum stacking. 
We explain our analysis to estimate properties of outflowing gas 
in Section \ref{sec:3}.
We provide our results of the outflow velocities 
in Section \ref{sec:4}, 
and discuss the redshift evolution of the outflow properties 
in Section \ref{sec:5}.
Section \ref{sec:6} summarizes our results.
We calculate stellar masses and SFRs by assuming
a \citet{Chabrier:2003} initial mass function (IMF).
We adopt a $\Lambda$CDM cosmology with 
$\Omega_\mathrm{M} = 0.27$, $\Omega_\mathrm{\Lambda} = 0.73$, 
$h = H_0/(100\ \mathrm{km\ s^{-1}\ Mpc^{-1}}) = 0.70$,
$n_s = 0.95$, and $\sigma_8 = 0.82$ throughout this paper, 
which are the same parameters as those used in \citet{Behroozi:2013}.
All transitions are referred to by their wavelengths in vacuum.
Magnitudes are in the AB system.

\section{DATA \& SAMPLE SELECTION}
\label{sec:2}
To study galactic outflows, 
we construct three samples at $z\sim0$--2. 
A $z\sim0$ sample is drawn from the SDSS Data Release 7 \citep[DR7;][]{Abazajian:2009}, 
a $z\sim1$ sample is drawn from the DEEP2 Data Release 4 \citep[DR4;][]{Newman:2013}, 
and a $z \sim 2$ sample is drawn from \citet{Erb:2006c}.
Each spectrum of galaxies in three samples does not have 
the signal-to-noise ratio (S/N) high enough for analysis of 
absorption lines.
We therefore produce high S/N composites by stacking the spectra.
In this section, we describe each sample and 
explain the method of the spectrum stacking.
Properties of the stacked spectra are listed in Table \ref{tb:1}.
We discuss the selection biases between three samples in Section \ref{sec:evolution}.

\input{tb1.tex}

\subsection{Galaxies at $z\sim0$}
\label{sec:sdss}
We select star-forming galaxies at $z\sim0$ from 
the SDSS DR7 \citep{Abazajian:2009} 
main galaxy sample \citep{Strauss:2002}.
The spectra of these galaxies have 
a mean spectral resolution of $R \sim 2000$,
a dispersion of $69\ \mathrm{km\ s^{-1}\ pixel^{-1}}$, and
a wavelength coverage spanning between 3800 and 9200 \AA.
These spectra are taken with fibers of a 3$\arcsec$ diameter, 
that are placed at the center of the galaxies.
The SDSS imaging data are taken through 
a set of $u$, $g$, $r$, $i$, and $z$ filters \citep{Fukugita:1996}
using a drift-scanning mosaic CCD camera \citep{Gunn:1998}.

Since our targets are active star-forming disk galaxies, 
we need some galaxy properties for sample selection.
The galaxy properties are mainly taken from the MPA/JHU galaxy catalog\footnote{
The MPA/JHU galaxy catalog is available at 
\url{http://www.mpa-garching.mpg.de/SDSS/DR7}}.
Systemic redshifts $z\sys$ are derived with observed spectra and 
a linear combination of the galaxy template spectra.
Because this measurement may be affected by blueshifted absorption that 
outflowing gas produce, 
we compare $z\sys$ in the MPA/JHU catalog with redshifts measured 
by fitting H$\alpha$ emission lines alone with a Gaussian function.
We find that the difference of the two redshifts
is typically $< 5\ \mathrm{km\ s^{-1}}$ and negligible.
Stellar masses $M_*$ are 
obtained by the fitting to the photometry \citep{Kauffmann:2003, Salim:2007}.
SFRs within the fiber (\fibsfr) are measured from 
the extinction-corrected $\mathrm{H\alpha}$ emission-line flux, 
and total SFR are estimated by applying aperture correction 
with the photometry outside the fiber 
\citep{Brinchmann:2004}.
For our study, stellar masses and SFRs are converted 
from a \citet{Kroupa:2001} IMF to a \citet{Chabrier:2003} IMF 
with a correction factor of 0.93.
SFR surface densities $\Sigma_\SFR$ are defined as {\fibsfr}/$\pi R^2$, 
where $R$ is the physical length corresponding to 
the SDSS 1.5$\arcsec$ aperture radius.
The MPA/JHU catalog also includes the emission- and absorption-line fluxes 
\citep[e.g., $\mathrm{H\alpha}$, $\mathrm{H\beta}$, 
{[\OIII]}, {[\NII]} and $D_n(4000)$;][]{Tremonti:2004}
and the photometric properties (e.g., five photometric magnitudes).
As a parameter to distinguish disk galaxies from ellipticals, 
we use the fracDeV \citep{Abazajian:2004}, which is the coefficient 
of the best linear fitting that is a combination of 
exponential and \citet{de-Vaucouleurs:1948} rules.
When the fracDeVs of galaxies are less/greater than 0.8, 
they are defined as disk/elliptical galaxies.
Because our targets are disk galaxies, 
we use the axial ratio of the exponential fitting.
Using Table 8 in \citet{Padilla:2008}, 
we calculate the inclinations, $i$, of our galaxies
from the $r$-band axial ratios and absolute magnitudes.

We use the similar criteria used by \citet{Chen:2010} 
to select the star-forming galaxies 
from the main galaxy sample \citep{Strauss:2002}, 
which contains the galaxies with 
extinction-corrected Petrosian $r$ magnitude in the range of $14.5 < r < 17.5$.
The values of $z\sys$ range from 0.05 to 0.18.
To select active star-forming disk galaxies, 
we apply the criteria of $D_n(4000)$ less than 1.5 and $r$-band fracDeV less than 0.8.
Active galactic nuclei (AGN) are excluded with the classification by \citet{Kauffmann:2003}.
We also exclude the galaxies 
whose SFRs or stellar masses are not derived in the MPA/JHU catalog.
Moreover, we set two additional selection criteria for our study.
First, we select the galaxies with 
$\Sigma_\SFR \geq 10^{-0.8}\ \mathrm{M_\odot\ yr^{-1}\ kpc^{-2}}$, 
which is above the empirical threshold 
$\Sigma_\SFR > 0.1\ \mathrm{M_\odot\ yr^{-1}\ kpc^{-2}}$ 
for local galaxies to launch outflows \citep{Heckman:2002}.
Second, we chose face-on galaxies whose inclinations are $i < 30^\circ$ 
because the typical opening angle of the outflows is $< 60^\circ$ for the SDSS galaxies 
\citep{Chen:2010}. 
There are 1321 galaxies that satisfy all of the selection criteria.
The blue dashed line in Figure \ref{fig:distribute} indicate 
the normalized distribution of stellar masses for this sample.
In order to make samples with similar stellar mass distribution at $z\sim0$--$2$, 
we randomly select the high-mass galaxies from the sample 
to match the distribution of the sample at $z\sim0$ to 
that of the sample at $z\sim1$ (Section \ref{sec:deep2}).
The final galaxy sample contains 802 galaxies.
We refer to the final sample as the \zz-sample.
The normalized distribution of stellar masses for the \zz-sample 
is shown in Figure \ref{fig:distribute} with the blue solid line.
The \zz-sample is plotted in Figure \ref{fig:ms} with the blue circles.
The median stellar mass of the \zz-sample is $\log(M_*/M_\odot)=10.46$.

We produce high S/N composites by stacking the spectra.
Regarding each individual spectrum, 
the wavelength is shifted to the rest-frame with $z\sys$, 
and the flux is normalized to 
the continuum around 
\NaID \lamlam5891.58, 5897.56.
Since bad pixels are identified as {\tt OR\_MASK} 
by the SDSS reduction pipeline, 
we exclude them in the same manner as \citet{Chen:2010}. 
These individual spectra are combined with an inverse-variance weighted mean.
We divide the \zz-sample into SFR bins 
where the composite spectra have S/N per pixel of 300 in the range of 6000--6050 \AA.

\subsection{Galaxies at $z\sim1$}
\label{sec:deep2}
For our galaxies at $z\sim1$, 
we use the DEEP2 DR4 \citep{Newman:2013} 
\footnote{The DEEP2 DR4 is available at 
\url{http://deep.ps.uci.edu/DR4/home.html}}. 
The survey is conducted with the DEIMOS \citep{Faber:2003} 
at the Keck II telescope. 
The DEEP2 survey targets galaxies with a magnitude limit 
of $18.5 < R_\mathrm{AB} < 24.1$
in four fields of the Northern Sky.
In three of the four fields, the DEEP2 survey preselect these galaxies with
$B$, $R$, and $I$ photometry taken with 12K camera 
at the Canada-France-Hawaii Telescope
to remove galaxies at $z < 0.7$ \citep{Coil:2004a}.
The spectra of these galaxies are taken with a 1200 lines/mm grating and
a 1\farcs0 slit. The spectral setting gives $R\sim5000$. 
The wavelength ranges from 6500 to 9100 \AA.
The public data are reduced with the DEEP2 DEIMOS Data pipeline 
\citep[the {\it spec2d} pipeline;][]{Newman:2013, Cooper:2012}.
In addition we execute IDL routines for the flux calibration
of the DEEP2 spectra 
\citep{Newman:2013}.
The IDL routines correct spectra for the overall throughput, 
chip-to-chip variations, and telluric contamination.
After these corrections, the IDL routines calibrate fluxes of the spectra with 
the $R$- and $I$-band photometry.
The flux calibration is accurate to 10\% or better.
Since the routines can not calibrate some spectra correctly, 
we exclude the spectra from our sample.

We use galaxy properties taken from the DEEP2 DR4 redshift catalog.
The catalog includes the absolute $B$-band magnitude $M_B$, 
($U-B$) color in the rest frame, 
systemic redshift $z\sys$, and object classification.
The last two parameters are determined 
by the {\it spec1d} redshift pipeline.
The {\it spec1d} pipeline finds the best value of $z\sys$ with galaxy-, QSO-, 
and star-template spectra 
to minimize $\chi^2$ of the data and the templates \citep{Newman:2013}.
The error of $z\sys$ is $\sim 16\ \mathrm{km\ s^{-1}}$.
We also use ($B-V$) color measured by C. N. A. Willmer (in private communication).

We use 
the \MgI $\lambda$2852.96 and  \MgII \lamlam2796.35, 2803.53 absorption lines 
in this study.
These lines fall in the spectral range of DEEP2 spectra of 
galaxies at $1.2 < z < 1.5$.
The \FeII \lamlam2586.65, 2600.17 lines are also useful for outflow studies 
because they are free from resonance scattering.
In the spectral range of the DEEP2 spectra, however, 
the \FeII lines are not available 
within the redshift range of the DEEP2 observation.

Since we need the spectra within which 
the \MgI and \MgII absorption lines fall, 
we firstly adopt the criteria used in \citet{Weiner:2009}.
\citet{Weiner:2009} select the spectra extending to 2788.7 \AA\ in the rest frame. 
To avoid the AGN/QSO contamination, 
\citet{Weiner:2009} exclude the galaxies at $z \geq 1.5$ or 
having the AGN classification of \citet{Newman:2013}.
In addition to these criteria of \citet{Weiner:2009}, 
we remove galaxies that are at the red sequence 
\citep{Willmer:2006,Martin:2012}.
We also remove the low-S/N galaxies 
that significantly affect the normalization procedure described below.
Our final sample contains 1337 galaxies at $1.2 \lesssim z\sys < 1.5$
with the median value $\langle z \rangle = 1.37$.
We refer to this sample as the \zo-sample.

\begin{figure}[t]
  \epsscale{1}
  \plotone{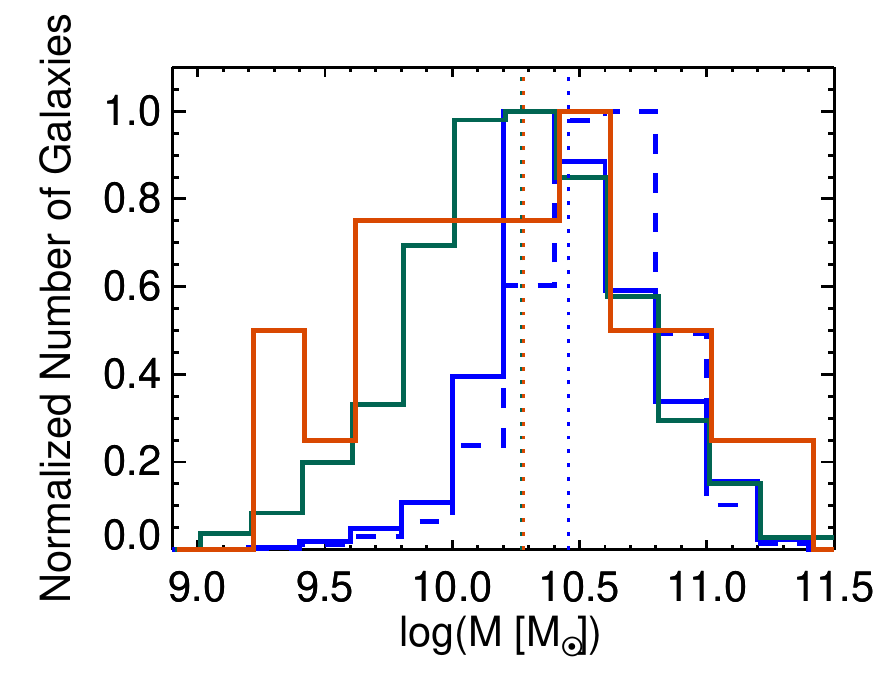}
  \caption{Normalized histograms of the stellar masses 
    for the three samples.
    The blue, green, and orange solid lines indicate 
    802 galaxies in the \zz-sample, 1337 galaxies in the \zo-sample, 
    and 25 galaxies in the \zt-sample, respectively.
    The blue dashed line denotes
    the SDSS galaxies that satisfy the selection criteria.
    The blue, green, and orange vertical doted lines show 
    the median stellar masses of the \zz-, \zo-, and \zt-samples.
    Each histogram is normalized to the maximum number of the bins.
  }
  \label{fig:distribute}
\end{figure}

We estimate stellar masses and SFRs from equations in the
literature.  Stellar masses are calculated from 
$M_B$, ($U-B$), and ($B-V$) colors 
by Equation (1) of \citet{Lin:2007}.
We rewrite the equation from Vega to AB magnitudes using the transformation in
\citet{Willmer:2006} and \citet{Blanton:2007a}:
\begin{equation}
  \begin{split}
    \log(M_*/M_\odot) =
    &- 0.4(M_B - 5.48) + 1.737(B - V) \\
    &+ 0.309(U - B) - 0.130(U - B)^2 \\
    &- 0.268z + 1.123.
  \end{split}
\end{equation}
The normalized distribution of stellar masses for the \zo-sample 
is shown in Figure \ref{fig:distribute} with the green line.
The median stellar mass of the \zo-sample is $\log(M_*/M_\odot)=10.24$.
SFRs are also calculated from $M_B$ and ($U-B$).
Using Equation (1) and Table 3 of
\citet{Mostek:2012}, we derive SFRs by
\begin{equation}
  \begin{split}
    \log\SFR = 
    &0.381 -0.424 (M_B+21) \\ 
    &- 2.925 (U-B) - 2.603 (U-B)^2.
  \end{split}
  \end{equation}
The quantities of the SFRs are calculated 
with a \citet{Salpeter:1955} IMF.
To obtain the quantities in a \citet{Chabrier:2003} IMF, 
we apply a correction factor of 0.62.
The \zo-sample is plotted in Figure \ref{fig:ms} with the green triangles.

Stacked spectra are produced in the same manner as in Section \ref{sec:sdss}, 
but with three procedures for the \zo-sample.
The first procedure is to convert wavelength from air to vacuum 
by the method of \citet{Ciddor:1996}.
The second one is to normalize the flux at the continuum measured 
in the wavelength range around \MgII
\lamlam2796.35, 2803.53.
The third one is to divide the \zo-sample into three subsamples in 
low, medium, and high SFR bins.

\subsection{Galaxies at $z\sim2$}
We use the sample of \citet{Erb:2006c} for our sample at $z\sim2$.
The sample consists of BX/BM \citep{Adelberger:2004, Steidel:2004} 
and MD \citep{Steidel:2003} galaxies, which are originally in 
the rest-frame UV-selected sample described by \citet{Steidel:2004}.
The rest-frame UV spectra of galaxies are taken with the LRIS at 
the Keck I telescope, and 
the near-infrared (NIR) spectra are mainly taken with NIRSPEC \citep{McLean:1998} 
at Keck II telescope.
Since our aim is to analyze rest-frame UV absorption lines, 
we obtain the raw data taken with the 400/3400 grism of LRIS-B 
from 2002 to 2003 (PI Steidel) through the Keck Observatory Archive\footnote
{KOA website: 
\url{http://www2.keck.hawaii.edu/koa/public/koa.php}} (KOA).
The spectral resolution is $R\sim800$.
The wavelength ranges from 3000 to 5500 \AA.

The data are reduced with the XIDL LowRedux\footnote
{The XIDL LowRedux is available at 
\url{http://www.ucolick.org/~xavier/LowRedux/}} pipeline.
Right ascension (R.A.) and declination (Decl.) of each galaxy are 
derived from the pixel values and the fits header of data 
with the program {\tt COORDINATES}\footnote
{{\tt COORDINATES} is available at 
\url{http://www2.keck.hawaii.edu/inst/lris/coordinates.html}}.
We find systematic errors of about 20$\arcsec$ in results of {\tt COORDINATES}.
However, since some galaxies are observed with one mask at the same time, 
we can correct R.A. and Decl. of them to be consistent with coordinates of 
\citet{Erb:2006c}.
As shown in \citet{Steidel:2010}, 
this sample contains galaxy--galaxy pairs;  
the circum-galactic medium around foreground galaxies 
give rise to absorption lines in the spectra of the background galaxies.
For this reason, we remove 6 background galaxies of the galaxy--galaxy pairs 
from our sample.
Finally, we obtain the spectra of 25 galaxies in the \cite{Erb:2006c} sample.
We refer to this sample as the \zt-sample.

\begin{figure}[t]
  \epsscale{1}
  \plotone{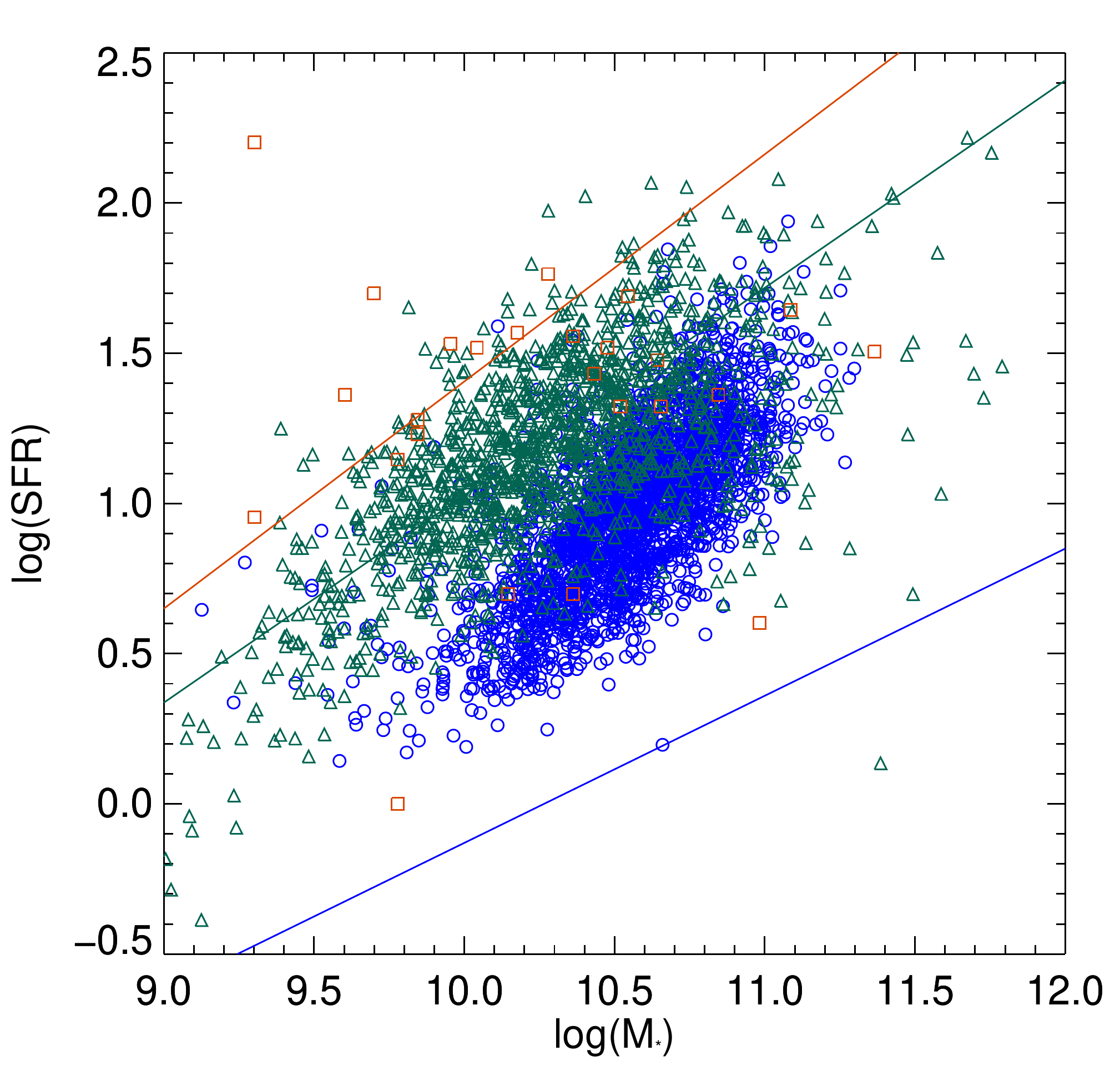}
  \caption{SFRs as a function of $M_*$ for our three samples.
    The blue circles, the green triangles, and the orange squares indicate 
    the galaxies of the \zz-, \zo-, and \zt-samples, respectively.    
    The intrinsic scatter of 0.2 dex \citep{Mostek:2012} 
    is included in the data points to SFRs of the z1-sample for the display purposes.
    The blue, green and orange solid lines are the star-forming main sequences 
    calcurated with Equation (28) of \citet{Speagle:2014a} 
    at $z\sim0$, $1$, and $2$, respectively.}
  \label{fig:ms}
\end{figure}

We take $z\sys$, stellar masses, and SFRs of the \zt-sample from \citet{Erb:2006c}.
The systemic redshifts $z\sys$ are determined by the H$\alpha$ emission line 
on the NIR spectra.
The typical rms error of $z\sys$ is $60\ \mathrm{km\ s^{-1}}$ \citep{Steidel:2010}.
Stellar masses and SFRs are derived from the SED fitting with 
the $U_n$, $G$, $R$, $J$, and $K$-band magnitudes.
The mid-infrared magnitudes taken with Infrared Array Camera  
on the Spitzer Space Telescope are also used, if they are available.
The normalized distribution of stellar masses for the \zt-sample 
is shown in Figure \ref{fig:distribute} with the orange line.
The \zt-sample is plotted in Figure \ref{fig:ms} with the orange squares.
The median stellar mass of the \zt-sample is $\log(M_*/M_\odot)=10.28$.

Stacked spectra are produced in the same manner as in Section \ref{sec:sdss}, 
but with four procedures for the \zt-sample: 
(1) to convert the wavelength from air to vacuum;
(2) to normalize the flux in the wavelength range between 1410 and 1460 \AA; 
(3) not to exclude bad pixels since they are not detected by the pipeline; and 
(4) not to divide the \zt-sample into subsamples.

\section{Analysis}
\label{sec:3}
We analyze metal absorption resonance lines in the galaxy spectra 
to study the outflow properties.
The absorption lines are \NaID \lamlam 5891.58, 5897.56 for the \zz-sample;
\MgI $\lambda$2852.96 and \MgII \lamlam2796.35, 2803.53 
for the \zo-sample; and 
\SiII $\lambda$1260, 
\CII $\lambda$1334.53, 
\SiII $\lambda$1527, and 
\CIV \lamlam1548.20, 1550.78 for the \zt-sample.
We assume that absorption profiles consist of three components: 
the intrinsic component composed of the stellar absorption and the nebular emission lines, 
the systemic component produced by the static gas in the interstellar medium (ISM) of the galaxies,
and the outflow component produced by the outflowing gas from the galaxies.
The stellar atmospheres of cool stars specifically give rise to 
the strong \NaID absorption, 
which impacts on the results of the outflow analysis \citep{Chen:2010}.
The stellar atmospheres also provide moderate absorption features of 
\MgI, \MgII, \SiII, \CII and \CIV
\citep{Rubin:2010, Coil:2011, Steidel:2016}.
Additionally, 
the static gas in the ISM produces absorption lines 
at the systemic velocity 
while the outflowing gas makes blueshifted absorption lines.
\citep[e.g.,][]{Martin:2005, Chen:2010, Rubin:2014}.
We follow the procedures of the analysis shown in \citet{Chen:2010}.
Below, we explain the procedures that are made in two steps.
First, we determine the stellar continuum of the stacked spectra 
for the intrinsic components of the absorption lines.
Second, we model the absorption lines with the stellar continuum 
and obtain the outflow components.

\begin{figure}[t!]
  \epsscale{1.0}
  \plotone{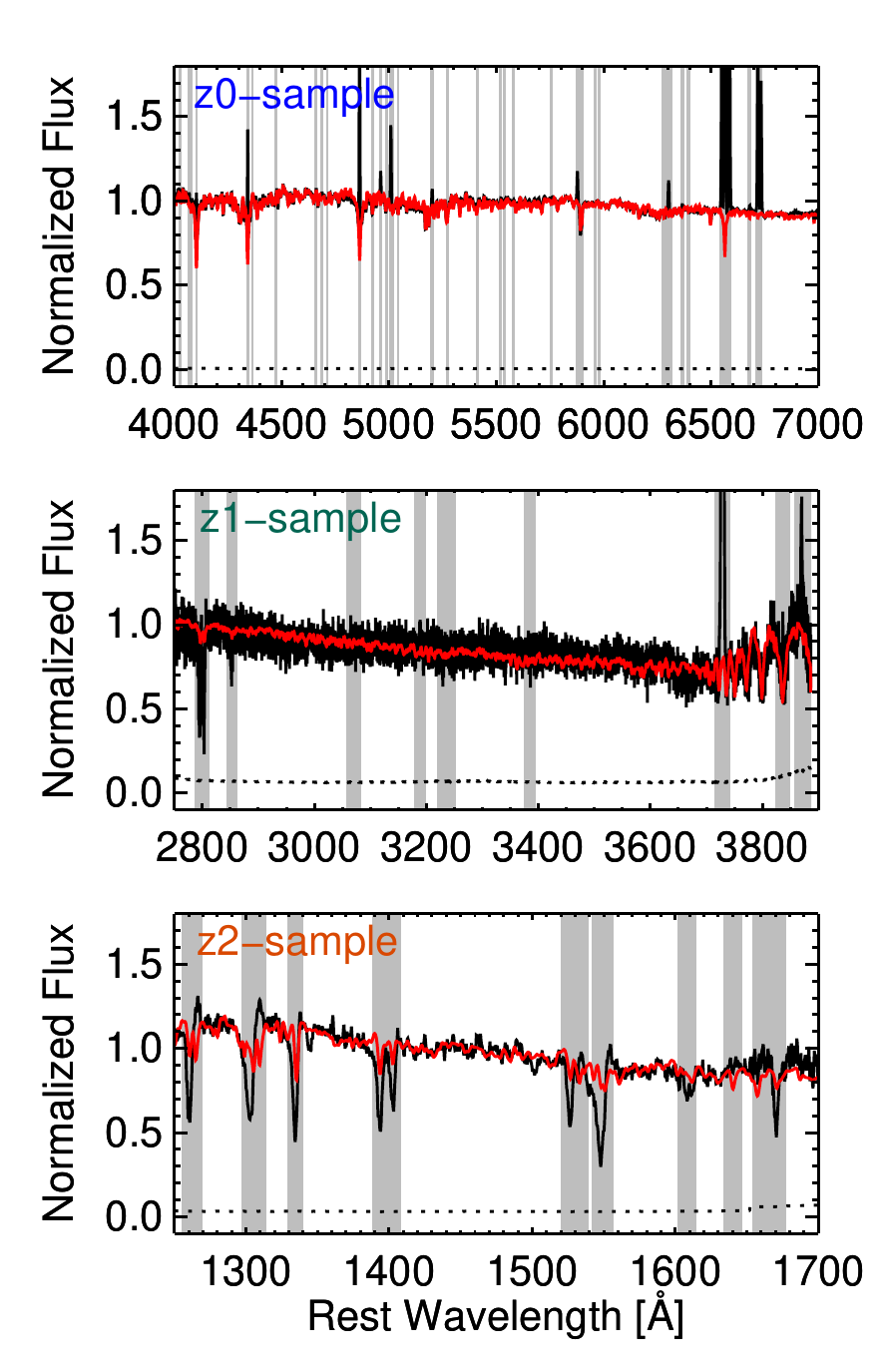}
  \caption{Examples of the stacked spectra (black line) and 
    the best-fitting continuum models (red line). 
    The spectra of the \zz-, \zo-, and \zt-samples are shown 
    from top to bottom.
    The wavelength range shown in this figure 
    is used for the stellar continuum fitting,
    except for the gray shading.
    The doted lines denote $1\sigma$ uncertainties of the spectra.}
  \label{fig:contfit}
\end{figure}

\begin{figure*}[!t]
  \epsscale{1}
  \plotone{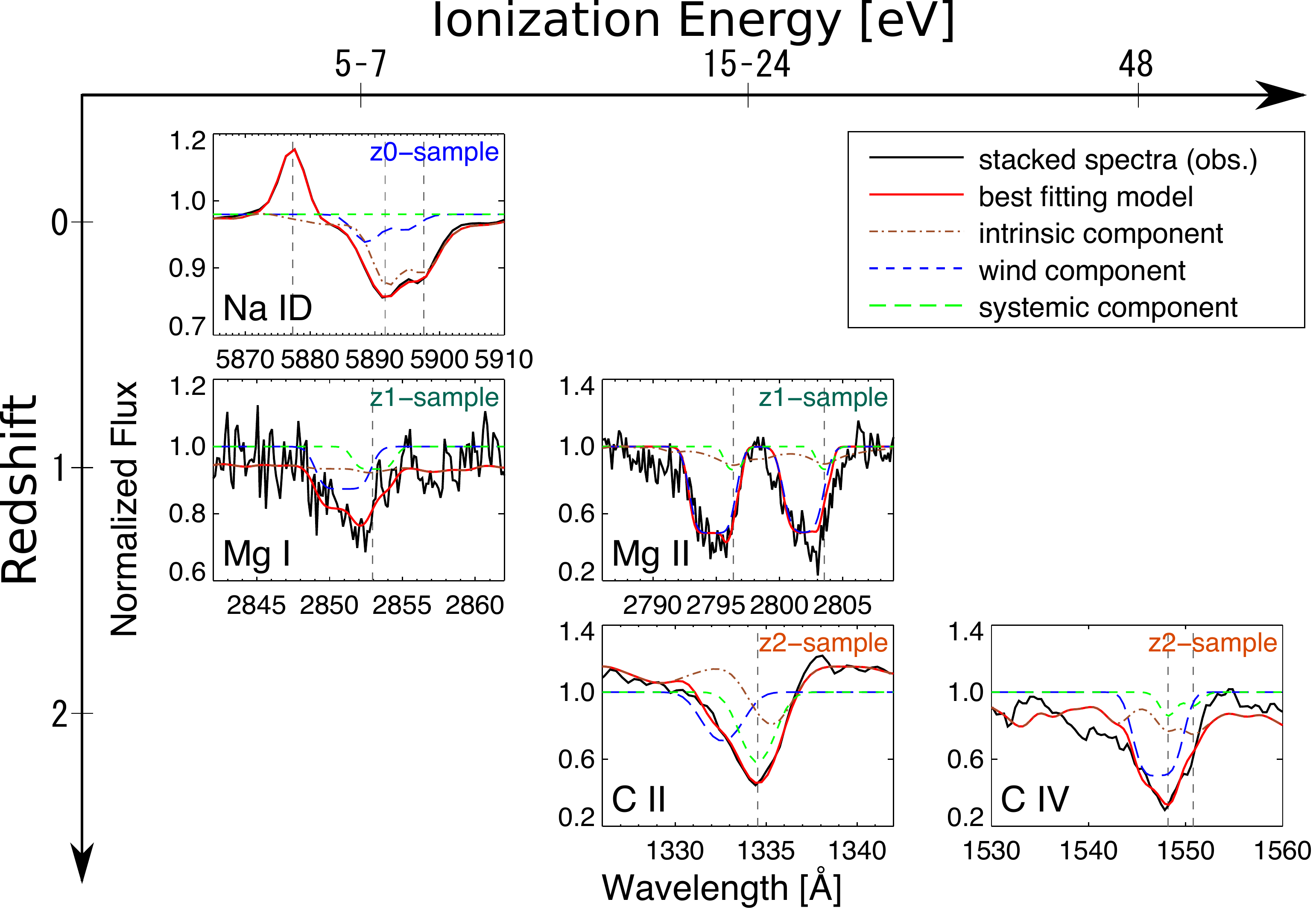}
  \caption{Examples of the stacked spectra around the absorption 
    lines (black line).
    The red lines represent the best fitting models.
    The dot-dashed brown, dashed blue, and long-dashed green indicate 
    the breakdowns of the lines for 
    the intrinsic, wind, and systemic components, 
    respectively.
    In this figure, the redshift (the ionization energy) 
    increase from top to bottom (from left to right).
    The ions of the absorption lines are written 
    at the bottom-left corner of the panels, 
    and the samples are written at the top-right corner.
    The vertical dashed gray lines denote 
    the rest-frame wavelengths of the absorption lines, 
    except for the line at 5877.29 \AA\ in the top-left panel, 
    which represents the wavelengths of \HeI emission.
  }
  \label{fig:windtable}
\end{figure*}

\subsection{Stellar Continuum Determination}
\label{sec:stellarcont}
We determine the stellar continuum with simple stellar population (SSP) models. 
For the \zz-sample, we adopt \citet[][BC03]{Bruzual:2003} SSP models, which
have a high spectral resolution in the wavelength of the SDSS spectra.
We use the 30 template spectra with 10 ages of 
0.005, 0.025, 0.1, 0.29, 0.64, 0.90, 1.4, 2.5, 5, and 11 Gyr, 
and three metallicities of $Z=0.004, 0.02$, and 0.05.
On the other hand, for the \zo- and \zt-samples, 
we adopt \citet{Maraston:2009} SSP models based on Salpeter IMF 
because the template spectra of \citetalias{Bruzual:2003} models have 
a low spectral resolution at wavelengths less than 3300 \AA.
\citet{Maraston:2009} models have a high spectral resolution of $R \sim 10000$ 
in the wavelength range of 1000--4700 \AA.
We use the 30 template spectra with 10 ages of 1, 5, 25, 50, 100, 
200, 300, 400, 650, and 900 Myr, 
and three metallicities of $Z=0.001, 0.01$, and 0.02.
For all samples, 
each template spectrum is convolved with the stellar velocity dispersion 
of each stacked spectrum.
We construct the best fitting models of the stacked spectra 
with a linear combination of the template spectra applying 
the starburst extinction curve of \citet{Calzetti:2000} 
by the {\tt IDL} routine {\tt MPFIT} \citep{Markwardt:2009}.
We make the linear combination of all 30 template spectra 
and fit to the \zz-sample composite spectra, which have high S/N.
We perform the same analysis of the \zo- and \zt-sample spectra, 
but with the linear combination of 10 template spectra of 
a fixed metallicity.
Since our main results are insensitive to differences of metallicity, 
we use $Z=0.01$ for \zo- and \zt-samples.
The rest-frame wavelength ranges used for fitting 
are 4000--7000 \AA\ for the \zz-sample,
2750--3500 \AA\ for the \zo-sample, 
and 1200--1600 \AA\ for the \zt-sample.
In the fitting we omit the wavelength ranges of 
all of the emission and absorption lines 
except for made by the stellar atmosphere.
Figure \ref{fig:contfit} shows 
the stacked spectra (black) and the best fitting models (red) of 
the \zz-, \zo-, and \zt-samples.

\subsection{Outflow Profile Estimate}
\label{sec:ope}
With the best fitting models of the stellar continuum, 
we estimate the systemic and outflow components of the absorption lines.
The normalized 
line intensity $I\obs(\lambda)$ is assumed to be expressed 
by three components: 
\begin{equation}
I\obs(\lambda) = I\intr(\lambda)\ I\sys(\lambda)\ I\out(\lambda), 
\label{eq:basic} 
\end{equation}
where $I\intr(\lambda)$ is the intrinsic component, and 
$I\sys(\lambda)$ and $I\out(\lambda)$ are the systemic and the outflow components 
whose continua are normalized to unity, respectively.
We use the stellar continuum given in Section \ref{sec:stellarcont} 
for $I\intr(\lambda)$, except for \NaID 
that are explained later in (ii).
We adopt a model made from a set of following two equations 
for each of the systemic and outflow components \citep{Rupke:2005a}.
In this model, the normalized line intensity $I(\lambda)$ is given by 
\begin{equation}
I(\lambda) = 1 - C_f+C_fe^{-\tau(\lambda)}, 
\label{eq:raditra}
\end{equation}
where $\tau(\lambda)$ is the optical depth, 
and $C_f$ is the covering factor.
Although $C_f$ may be a function of wavelength \citep{Martin:2009}, 
we assume that $C_f$ is independent of wavelength.
Under the curve-of-growth assumption, 
the optical depth is written with a Gaussian function as
\begin{equation}
\tau(\lambda) = \tau_0 e^{-(\lambda-\lambda_0)^2/(\lambda_0 b_D/c)^2}, 
\label{eq:tau}
\end{equation}
where $c$ is the speed of light, 
$\tau_0$ is the optical depth at the central wavelength $\lambda_0$ 
of the line, 
and $b_D$ is the Doppler parameter in units of speed.
There are 4 parameters 
for each of $I\out(\lambda)$ 
(i.e., $\lambda\oout{0}$, $\tau\oout{0}$, $C\oout{f}$, and $b\oout{D}$)
and $I\sys(\lambda)$ 
(i.e., $\lambda\osys{0}$, $\tau\osys{0}$, $C\osys{f}$, and $b\osys{D}$).
Because the central wavelength $\lambda\osys{0}$ of $I\sys(\lambda)$ is fixed 
to the rest-frame wavelength $\lambda\rest$.
Hence, 
the model of $I\obs(\lambda)$ includes 7 ($=4+3$) free parameters in total.

Equation (\ref{eq:basic}) is modified in the following two cases.
(i) The first case is applied for the doublet lines
(\NaID, \MgII, and \CIV).
We define the optical depth of the blue and red lines of the doublet 
as $\tau_\mathrm{B}(\lambda)$ and $\tau_\mathrm{R}(\lambda)$
whose central wavelengths are $\lambda_\mathrm{0,B}$ and $\lambda_\mathrm{0,R}$, respectively.
The total optical depth is written as  
$\tau(\lambda) = \tau_\mathrm{B}(\lambda)+\tau_\mathrm{R}(\lambda)$.
Because the blue lines have oscillator strengths twice higher than 
the red lines for all of the doublet lines, 
the central optical depths of the blue lines $\tau_{0,\mathrm{B}}$ 
are related to those of the red lines $\tau_{0,\mathrm{R}}$ by 
$\tau_{0,\mathrm{R}} = \tau_{0,\mathrm{B}}/2$.
The ratio of the central wavelengths 
$\lambda_\mathrm{0,B}/\lambda_\mathrm{0,R}$ of the outflow component is fixed 
to the rest-frame doublet wavelengths ratio.
We assume that $b_D$ and $C_f$ for the blue and red lines
are the same since these lines should arise from the same gas clumps.
The number of free parameters therefore remains unchanged.
(ii) The second case is applied for \NaID, 
which has a neighboring emission line \HeI $\lambda$5877.29.
We model the emission line profile $I\emi(\lambda)$ of \HeI 
using a Gaussian function with additional 2 free parameters. 
In this case, 
the intrinsic component of Equation (\ref{eq:basic}) is expressed as 
\begin{equation}
  \label{eq:emi}
I\intr(\lambda) = I\cont(\lambda) + I\emi(\lambda), 
\end{equation}
where $I\cont(\lambda)$ is the stellar continuum given 
in Section \ref{sec:stellarcont}.
For \NaID, there are the total of 9 ($=7+2$) free parameters 
in $I\obs(\lambda)$ that are composed of 7 free parameters of 
Equations (\ref{eq:raditra})--(\ref{eq:tau}) and 
the 2 parameters of the Gaussian function $I\emi(\lambda)$.

We carry out the model fitting and 
obtain the best-fit model of each absorption line 
with {\tt MPFIT}.
We place three constrains of the parameters for the fitting.
First,
since \citet{Martin:2009} show that 
covering factors of \NaID and \MgI are smaller than those of \MgII, 
we limit the range of $C\oout{f}$ to 
$0 \leq C\oout{f} \leq 0.2$ ($0 \leq C\oout{f} \leq 1$)
for \NaID and \MgI (the other absorption lines).
Second,
we fix $b\oout{D}$ of the \zt-sample at $130\mathrm{km\ s^{-1}}$ that are 
the average value of the \zo-sample, 
and we constrain $b\osys{D}$ of the \zt-sample 
at $\leq 150\mathrm{km\ s^{-1}}$.
Third, 
we set the upper limit of $\tau\oout{0}$ 
so that the column densities of hydrogen $N(\eH)$ 
calculated with Equations (\ref{eq:mlf4}) and (\ref{eq:nh}) 
are as much as $\log N(\eH) \leq 21.5$ 
\citep{Rupke:2005a, Martin:2006, Rubin:2014}.
Figure \ref{fig:windtable} shows the examples of the fitting results.
Our procedure provides the reasonable results, 
and detect the blueshifted outflow components 
that are shown with the dashed blue lines in Figure \ref{fig:windtable}.

Figure \ref{fig:windtable} indicates that 
the best-fit \NaID profile is composed of 
the large intrinsic and small outflow components.
Since the intrinsic component is estimated 
with the \citetalias{Bruzual:2003} SSP model, 
a different SSP model may systematically affect our results.
To evaluate the systematics 
we analyze the Na ID lines with another SSP model, 
\citet[][MS11]{Maraston:2011a} based on MILES.
The central and maximum outflow velocities 
estimated with \citetalias{Maraston:2011a} are 
very similar to those estimated with \citetalias{Bruzual:2003}; 
their differences are only 10--30 $\mathrm{km\ s^{-1}}$.
This is 
because the Na ID stellar absorption line of \citetalias{Maraston:2011a} 
is shallower than that of \citetalias{Bruzual:2003}.
This result is consistent with the appendix in \citet{Chen:2010}, 
which evaluate the systematics that different SSP models give.
\footnote{
As shown in Section \ref{sec:evolution}, 
a decrease in outflow velocities at $z\sim0$ strengthens 
the redshift evolution of outflow velocities from $z\sim0$ to $2$.
Thus systematics given by SSP models do not change our conclusion.
}

We note that 
there exists a plausible shallow absorption around 1540 \AA\ 
in the bottom-right panel of Figure \ref{fig:windtable}.
This shallow absorption may arise from stellar winds of young stars that 
broaden the \CIV absorption line \citep{Schwartz:2004, Schwartz:2006a}.
Although the outflow components are determined by a deep \CIV absorption line, 
the shallow absorption may produce 
a systematic uncertainty of the outflow velocities.
\citet{Du:2016a} reproduce the shallow absorption with SSP models 
from \citet{Leitherer:2010a} that includes 
the predicted effects of the stellar wind.
Because we do not use the \CIV absorption line for discussion, 
the results of \CIV are not relevant to our main scientific results.

\section{RESULTS}
\label{sec:4}
\subsection{Outflow Velocity}
\label{sec:outflowvelocity}
We define the velocity of the outflowing gas in two ways: 
the central outflow velocity $v\out = c(\lambda\rest-\lambda\oout{0})/\lambda\rest$ and 
the maximum outflow velocity 
$v\maxi = c(\lambda\rest-\lambda\maxi)/\lambda\rest$.
The maximum wavelength $\lambda\maxi$ is defined as 
\begin{equation}
\label{eq:lambdamax}
\lambda\maxi = \lambda\oout{0} - \lambda\oout{0}\frac{b\oout{D}}{c}
\sqrt{-\ln\left(\frac{1}{\tau\oout{0}}\ln\frac{1}{0.9}\right)}, 
\end{equation}
at which the blue side of the outflow component reaches 90 \% of the flux 
from $1 - C_f$ to unity, i.e., $I\out(\lambda\maxi) = 0.9(1-C_f)$.
The central outflow velocity $v\out$ is 
the central velocity of the outflowing gas, 
which represents the bulk motion of the gas.
In contrast, 
the maximum outflow velocity $v\maxi$ reflects the gas motion 
at the largest radii of the outflows, based on the simple scenario 
that the outflowing gas is accelerated towards the outside of the halo
\citep{Martin:2009}.
Therefore, $v\maxi$ is the indicator of 
whether the outflowing gas can escape the galactic halo to the inter-galactic medium (IGM).

\begin{figure}[t]
  \epsscale{1.0}
  \plotone{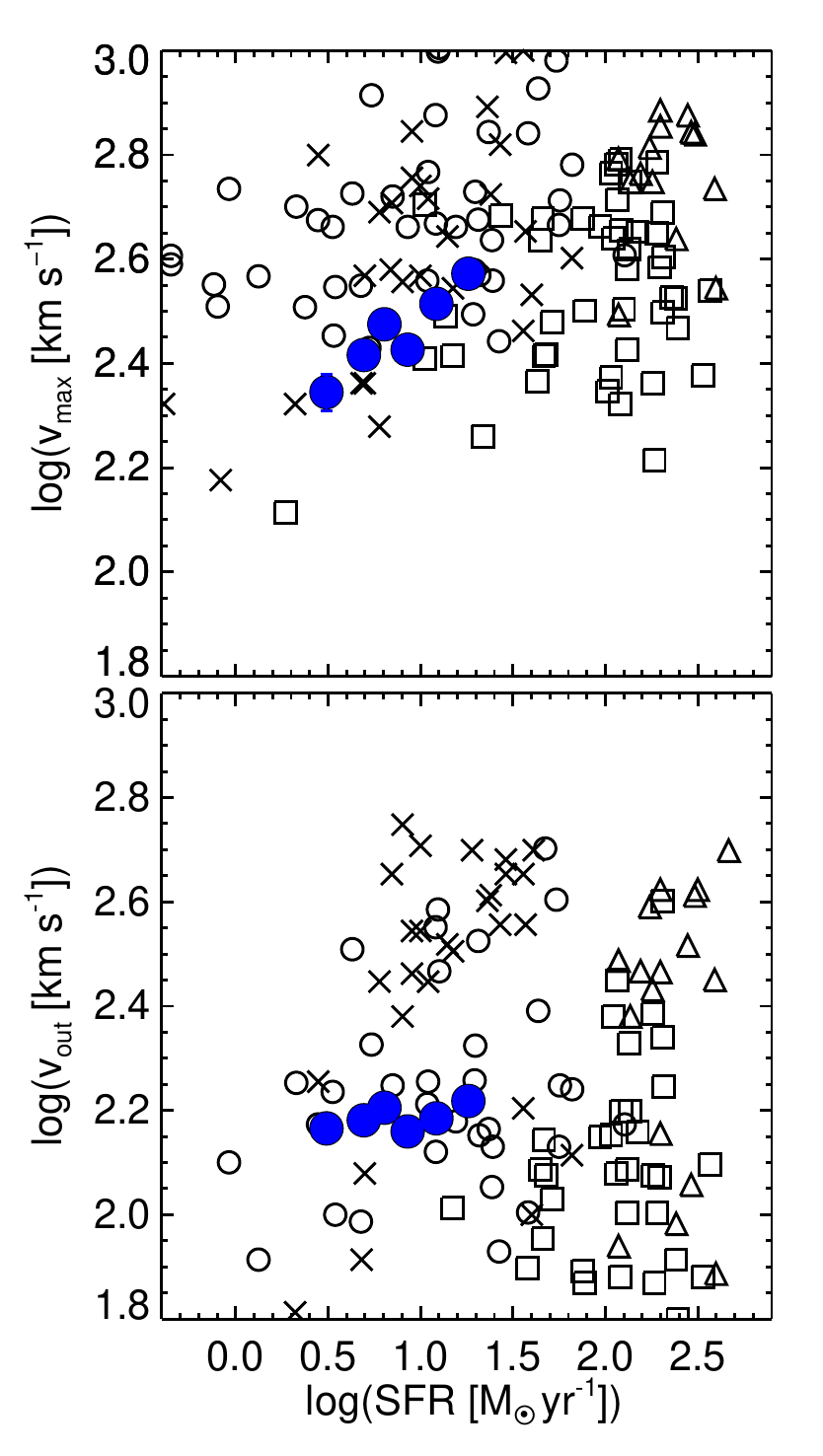}
  \caption{The outflow velocities at $z\sim0$.
    Top: the maximum outflow velocities (blue filled circle) of the \zz-sample 
    as a function of SFRs.
    Error bars denote the 1$\sigma$ fitting error.
    The open symbols show the maximum outflow velocities 
    of local galaxies in the literature: 
    \citet[cross]{Heckman:2015}, \citet[triangle]{Martin:2005}, 
    \citet[square]{Rupke:2005b}, and \citet[circle]{Chisholm:2015a}.
    Bottom: the central outflow velocities of the \zz-sample.
    The symbols are the same as those in the top panel of this figure.
  }
  \label{fig:vwind_sdss}
\end{figure}
\begin{figure}[t]
  \epsscale{1.0}
  \plotone{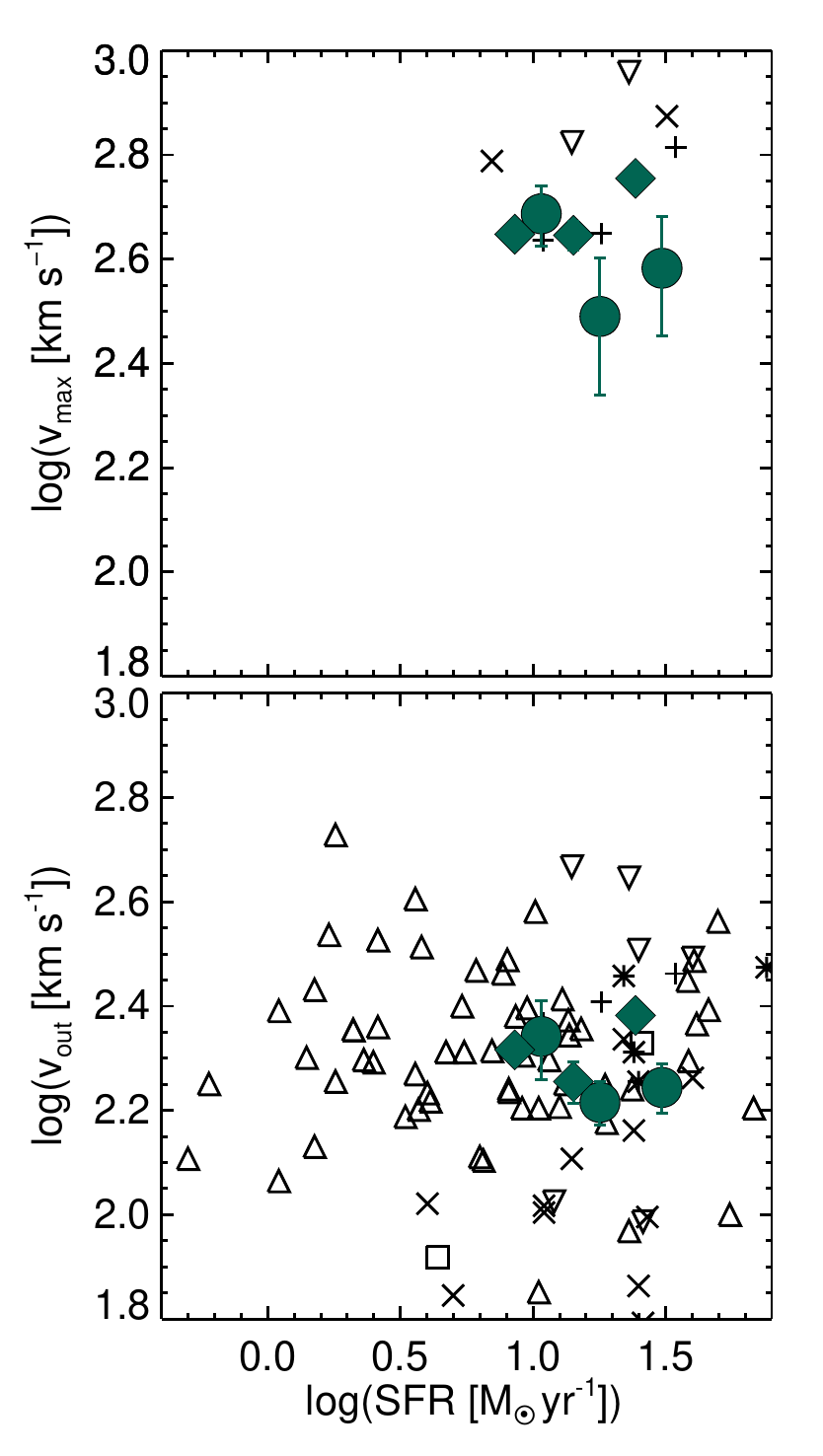}
  \caption{Same as Figure \ref{fig:vwind_deep}, 
    but for the outflow velocities at $z\sim1$.
    Top: the maximum outflow velocities of the \zo-sample for \MgI and \MgII
    (green filled circle and diamond, respectively) 
    as a function of SFRs..
    Error bars denote the 1$\sigma$ fitting errors.
    The open symbols show the maximum outflow velocities of 
    galaxies at $z\sim0.5$--$1$ in the literature: 
    \citet[cross]{Kornei:2012}, 
    \citet[triangle]{Rubin:2014}, \citet[square]{Bordoloi:2014b}, 
    \citet[plus]{Weiner:2009}, \citet[upside-down triangle]{Martin:2012}, 
    and \citet[asterisk]{Du:2016a}.
    Bottom: the central outflow velocities of the \zo-sample.
    The symbols are the same as those in the top panel of this figure.
    In both panels, the green diamonds are 
    offset in SFR by 0.1 dex for clarity.
  }
  \label{fig:vwind_deep}
\end{figure}
\begin{figure}[t]
  \epsscale{1.0}
  \plotone{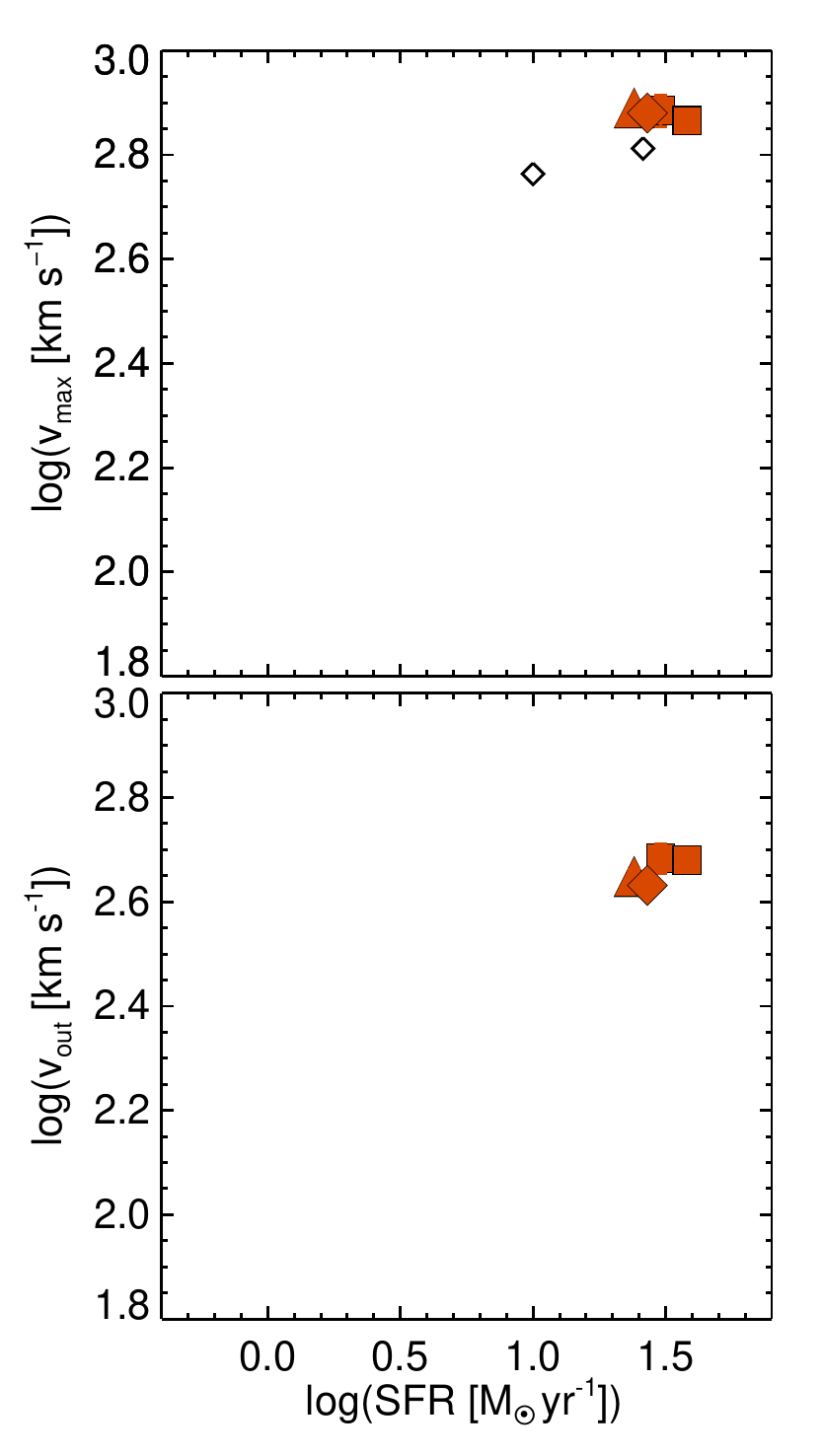}
  \caption{Same as Figure \ref{fig:vwind_sdss}, 
    but for the central outflow velocities at $z\sim2$.
    The orange diamond, triangle, and squares indicate 
    the central outflow velocities of the \zt-sample for 
    \CII, \CIV, and \SiII respectively.
    Error bars denote the 1$\sigma$ fitting errors.
    The open diamonds show the maximum outflow velocities of 
    galaxies in \citet{Erb:2012}.
    The orange triangle is offset in SFR by 0.1 dex for clarity.}
  \label{fig:vwind_lris}
\end{figure}

Figures \ref{fig:vwind_sdss}--\ref{fig:vwind_lris} 
show $v\maxi$ and $v\out$ as a function of SFRs.
In Figure \ref{fig:vwind_sdss}, 
$v\maxi$ and $v\out$ of the \zz-sample (shown with the blue circles) are 
as high as those of galaxies at $z\sim0$ in the literature (shown with the open symbols).
We perform a power-law fitting to $v\maxi$ and $v\out$ 
with the form of $V = V_1\, \SFR^{\alpha}$.
The best-fit parameters are 
$V_1=\vmZz\pm\vmZzerr$ and $\alpha = \amZz\pm\amZzerr$ for $V=v\maxi$; and 
$V_1=\voZz\pm\voZzerr$ and $\alpha = \aoZz\pm\aoZzerr$ for $V=v\out$.
The parameter $\alpha$ for $v\maxi$ shows a significant correlation 
between $v\maxi$ and SFR.
Our measurement of $\alpha = \amZz\pm\amZzerr$ for $v\maxi$ is consistent with 
the results of \citet{Martin:2005} and \citet{Martin:2012} 
who claim that $v\maxi$ has a steep slope of $\alpha = 0.35\pm0.06$.
On the other hand, 
the parameter $\alpha$ for $v\out$ suggests no correlation with SFR, 
which is supported by previous studies \citep{Chen:2010,Martin:2012}.
 
Figure \ref{fig:vwind_deep} compares 
$v\maxi$ and $v\out$ of the \zo-sample 
with those of galaxies at $z\sim1$ in the literature.
The \zo-sample shown with the green symbols is 
in good agreement with the literature.
Particularly, the values of $v\maxi$ is 
consistent with those in \citet{Weiner:2009}, 
who use the DEEP2 DR3 spectra.
We also find the weak positive scaling relation between $v\out$ and SFR.
In the literature, 
whereas $v\out$ of individual galaxies show no correlation 
\citep{Kornei:2012, Martin:2012, Rubin:2014}, 
those of stacked galaxies show the positive correlation 
\citep{Weiner:2009, Bordoloi:2014b}.
The positive scaling relation in Figure \ref{fig:vwind_deep} (the green symbols)
are consistent with the previous results of the stacked galaxies.

Figure \ref{fig:vwind_lris} shows 
the outflow velocities estimated with the \CII, \SiII, and \CIV lines of the \zt-sample.
Unlike the \CII and \CIV lines, 
the \SiII $\lambda$1260 and $\lambda$1527 lines have 
their associated \SiIIe $\lambda$1265 and $\lambda$1533 
fluorescent emission lines, respectively.
For this reason, 
emission filling of the \SiII absorption lines 
may be weaker than that of the \CII and \CIV lines.
Figure \ref{fig:vwind_lris} illustrates that 
this difference of the lines do not change
the outflow velocities in our analysis.
In the remainder of this paper, we use the \CII line to 
estimate outflow parameters of the \zt-sample.

Figure \ref{fig:vwind_lris} compares the \zt-sample and the literature at 
$z\sim2$.
\citet{Steidel:2010} study the outflow velocities ($\Delta v_\mathrm{IS}$)
of the UV-selected galaxies at $z\sim2$ 
with the sample drawn from \cite{Erb:2006c}.
The mean outflow velocity 
($\langle \Delta v_\mathrm{IS} \rangle = -164 \pm 16\ \mathrm{km\ s^{-1}}$)
is lower than $v\out$ of the \zt-sample (shown with the orange symbols).
This arises from different methods to measure the velocities: 
the outflow velocity become lower with the one-component fitting 
than with the two-component fitting.
Therefore, it is reasonable that $\Delta v_\mathrm{IS}$ is lower than 
$v\out$, even though the \zt-sample is a subsample drawn from \citet{Erb:2006c}.
\citet{Steidel:2010} also find that 
the scaling relation of central outflow velocities at $z\gtrsim2$ 
is flatter than that at $z<2$, but 
we can not mention the scaling relation of the \zt-sample due to 
the small sample size.

\subsection{Evolution of Outflow Velocity}
\label{sec:evolution}
\begin{figure}[!t]
  \epsscale{1.0}
  \plotone{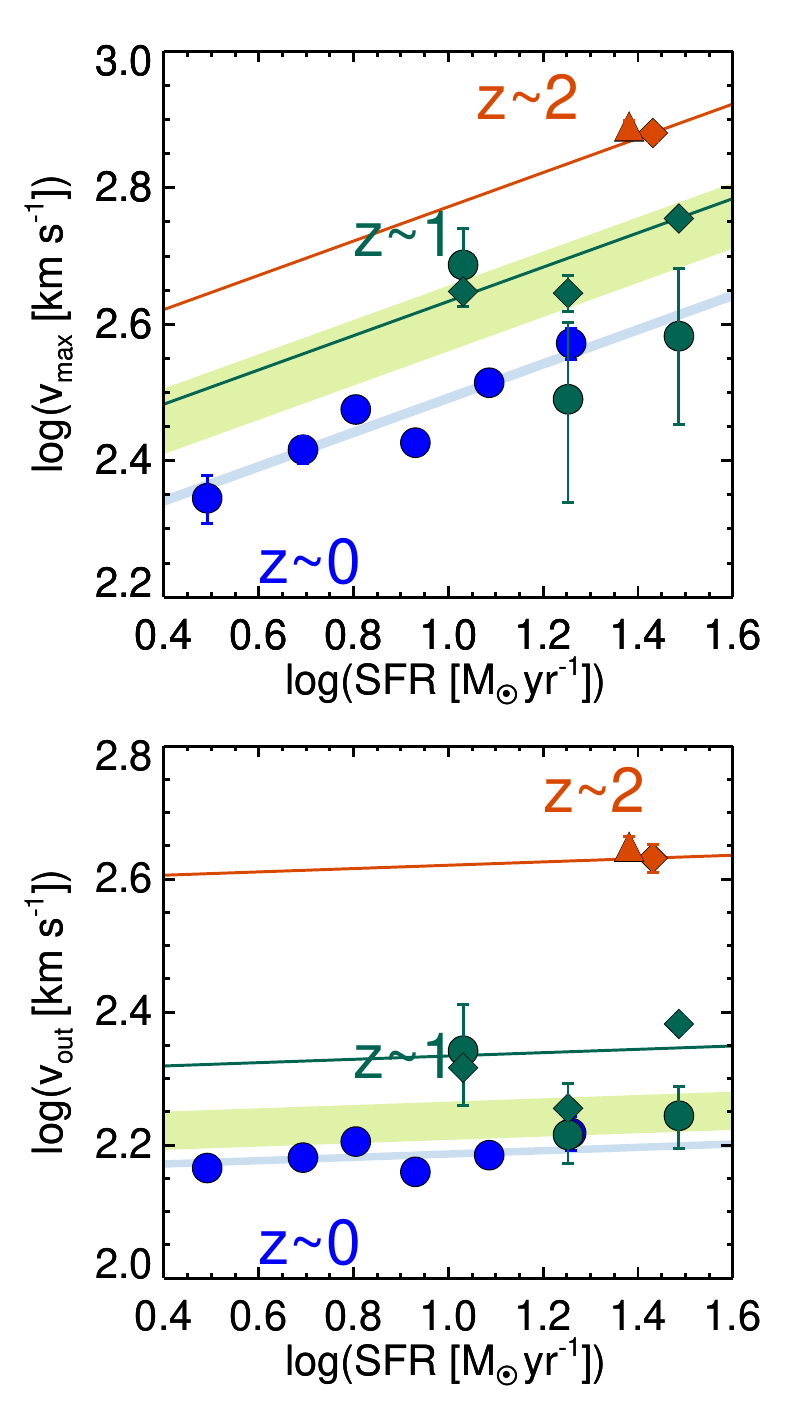}
  \caption{Same as Figure \ref{fig:vwind_sdss}, but for 
    the \zz- (blue symbol), \zo- (green symbol), and \zt- (red symbol) 
    samples.
    Top: the maximum outflow velocities as a function of SFRs.
    Each symbol corresponds to the elements of the absorption lines: 
    \NaID (blue circle), \MgI (green circle), 
    \MgII (green diamond), \CII (orange diamond), and \CIV (orange triangle).
    The circles, diamonds, and triangle indicate the velocities of elements, 
    which have the low (5--7 eV), medium (15--24 eV), and high (48 eV)
    ionization energy, respectively.
    Error bars denote the $1\sigma$ fitting errors.
    The light blue and green shades describe the result of the power-law fitting 
    to \NaID and \MgI, respectively, with vertical $1\sigma$ fitting error range.
    The green and orange lines denote the best-fitting power-law function 
    of \MgII and \CII, respectively.
    Bottom: the central outflow velocities as a function of SFRs.
    The symbols are the same as those in the top panel of this figure.
    In both panels, the orange triangles are
    offset in SFR by 0.05 dex for clarity.
}
  \label{fig:vwind}
\end{figure}
We show $v\maxi$ measurements 
of our \zz-, \zo-, and \zt-samples as a function of SFRs 
with the blue, green, and orange symbols, respectively, 
in the top panel of Figure \ref{fig:vwind}.
The bottom panel is the same as the top panel, but for $v\out$.
Figure \ref{fig:vwind} indicates 
the increasing trend of the outflow velocity with increasing redshift.
\citet{Martin:2009} and \citet{Chisholm:2016a} indicate that 
the outflow velocity depends on the depths of the absorption lines 
whereas \citet{Tanner:2016a} show that 
the outflow velocity depends on 
ionization energy (IE) of the ions used for velocity measurements.
For these reasons, 
we compare the outflow velocities of the absorption lines that have 
the similar depths and IE,  
which is presented in Figure \ref{fig:windtable}.
The details of our comparisons are explained below.

To compare the \zz- with \zo-samples, 
we use $v\maxi$ and $v\out$ computed from 
\NaID (IE $\simeq 5.1$ eV) and \MgI (IE $\simeq 7.6$ eV) absorption lines, 
respectively, 
which are depicted with the circles in Figure \ref{fig:vwind}.
In Section \ref{sec:outflowvelocity}, 
we obtain the best-fit parameter sets of 
the scaling relation $V=V_1\,\SFR^{\alpha}$ for \NaID of the \zz-sample: 
$V_1 = \vmZz\pm\vmZzerr\ (\voZz\pm\voZzerr)$ and 
$\alpha = \amZz\pm\amZzerr\ (\aoZz\pm\aoZzerr)$ for $V = v\maxi$ ($v\out$).
We perform a power-law fitting to $v\maxi$ ($v\out$) for 
\NaID of the \zz-sample and \MgI of the \zo-sample 
with the slope fixed at $\alpha = \amZz\ (\aoZz)$.
The best-fit parameter sets are 
$V_1=\vmZz\pm\vmZzerrfix\ (\voZz\pm\voZzerrfix)$ and 
$V_1=\vmmgi\pm\vmmgierr\ (\vomgi\pm\vomgierr)$ 
for \NaID and \MgI, respectively.
The blue and green shades in Figure \ref{fig:vwind} illustrate 
the best-fitting relations of \NaID at $z\sim0$ and \MgI at $z\sim1$, respectively.
The widths of the shades represent the 1$\sigma$ fitting error ranges.
Figure \ref{fig:vwind} indicates that 
$v\maxi$ ($v\out$) at $z\sim1$ is significantly higher than the one at $z\sim0$.

To compare the \zo- with \zt-samples, 
we use $v\maxi$ and $v\out$ computed from 
\MgII (IE $\simeq 15$ eV) and \CII (IE $\simeq 24$ eV) absorption lines, respectively, 
which are depicted with the diamonds in Figure \ref{fig:vwind}.
In the same manner as \MgI of the \zo-sample,
we fit $v\maxi$ ($v\out$) of \MgII by a power-law function 
with a slope fixed at $\alpha=\amZz\ (\aoZz)$.
We obtain the best-fit parameters of 
$V_1=\vmmgii\pm\vmmgiierr\ (\vomgii\pm\vomgiierr)$ for $V = v\maxi$ ($v\out$).
The green line in Figure \ref{fig:vwind} illustrate 
the best-fitting relations of \MgII at $z\sim1$.
For comparison, 
the orange line show the line with $\alpha=\amZz\ (\aoZz)$ 
through the orange diamond.
Figure \ref{fig:vwind} suggests that 
$v\maxi$ ($v\out$) at $z\sim2$ is significantly higher than the one at $z\sim1$.

We note that Figure \ref{fig:vwind} illustrates a large difference of $v\out$ 
between \MgII and \CII.
Since 
the systemic component in the \CII line is
larger than that in the \MgII line (Figure \ref{fig:windtable}), 
this large systemic component may generate the high $v\out$ value of \CII.
To evaluate outflow velocities with a small systemic component, 
we fit the \CII line with $C\osys{f} = 0.1$, 
which is the median of the best-fit $C\osys{f}$ values at $z \sim 0$--$1$, 
and without the constraints on $b\oout{D}$.
The best-fit $v\maxi$ $(\simeq 719 \pm 39\ \mathrm{km\ s^{-1}})$ is 
consistent with the values estimated without a constraint to $C\osys{f}$.
Thus the $v\maxi$ value is not affected by $C\osys{f}$.
Our conclusions do not change.
On the other hand, 
$v\out$ becomes small down to 
$\simeq 208 \pm 30\ \mathrm{km\ s^{-1}}$.
We think that the results of $v\out$ are easily affected by the systemic component and less reliable than those of $v\maxi$.

In summary, we find that $v\maxi$ and $v\out$ increase from $z\sim0$ to 2.
Although there is 
some implication of the redshift evolution of the outflow velocities 
\citep{Du:2016a, Rupke:2005b}, 
this is the first time to identify the clear trend of the redshift evolution 
of the outflow velocities.

Here we discuss the effects of the selection biases.
There are three sources of possible systematics that 
are included in our analysis.
The first is the selection criterion of the SFR surface density $\Sigma_\SFR$ 
in the \zz-sample.
In Section \ref{sec:sdss}, 
we select the galaxies with the criterion of $\Sigma_\SFR$ 
larger than $10^{-0.8}\ \mathrm{M_\odot\ yr^{-1}\ kpc^{-2}}$, 
but we do not apply this criterion for the \zo- and \zt-samples.
A large fraction of the \zo-sample meets the criterion of $\Sigma_\SFR$ 
because the galaxies of the \zo-sample have 
the median SFR of $\sim 7\ \mathrm{M_\odot\ yr^{-1}}$ and 
the median Petrosian radius of $5.2\ \mathrm{kpc}$ 
estimated from photometry of some galaxies 
taken with Hubble Space Telescope/Advanced Camera for Surveys 
\citep{Weiner:2009}.
All of our \zt-sample also meet the criterion of $\Sigma_\SFR$ 
\citep{Erb:2006c}.
The second is the selection criterion of the inclination $i$ in our \zz-sample.
We select the galaxies with $i < 30^\circ$, which is less than 
$60^\circ$ that is the typical outflow opening angle of the SDSS galaxies
\citep{Chen:2010}.
This criterion is likely not needed for the \zo- and \zt-samples 
because it is reported that 
the outflows of the galaxies at $z\sim1$--2 is more spherical than 
those at $z\sim0$ \citep{Weiner:2009,Martin:2012,Rubin:2014}.
In addition, the galaxies under these criteria of $\Sigma_\SFR$ and $i$ should 
decrease the $v\maxi$ and $v\out$ of the \zz-sample, 
indicating the redshift evolution of the outflow velocities more clearly.
The third is the differences of instrumental resolutions.
There is a possibility that 
low spectral resolutions may systematically increase the values of $v\maxi$.
We convolve the highest resolution ($R\sim5000$) spectra of \zo-sample 
with SDSS ($R\sim2000$) and LRIS ($R\sim800$) spectral resolutions and 
compare the $v\maxi$ values of the original and  the convolved spectra.
In this way, we confirm that 
the systematics of the different spectral resolution is negligible in our results.

\section{DISCUSSION}
\label{sec:5}
\subsection{Physical Origins of $v\maxi$ Evolution}
In Section \ref{sec:evolution}, 
we find that the outflow velocities increase from $z\sim 0$ to $2$.
The power-law fitting to the results of Figure \ref{fig:vwind} gives 
\begin{equation}
  \label{eq:po1-0}
  v\maxi \propto (1+z)^{\vz\pm\vzerr}
\end{equation}
at a fixed SFR.
This increasing trend of the outflow velocities is predicted by 
\citet{Barai:2015}, who carry out simulations with MUPPI 
and find that the outflow velocities increase from $z\sim0.8$ to $3.0$.
We cannot quantitatively compare our results with 
those of the simulation 
because of the different definitions of the outflow velocity.

The redshift evolution of $v\maxi$ is interpreted by 
an increase in $\Sigma_\SFR$ from $z\sim0$ to 2. 
\citet{Shibuya:2015} show that 
effective radii of galaxies decrease with increasing redshift by $r \propto (1+z)^{-1}$
at a fixed stellar mass. 
On the assumption that 
a projected surface area of a disk galaxy is proportional to $r^2$, 
$\Sigma_\SFR$ increases with increasing redshift by 
$\Sigma_\SFR \propto (1+z)^2$ at a fixed stellar mass for a given SFR.
Assuming the relation of $v\maxi \propto \Sigma_\SFR^{1/3}$ at $z=0$ 
found by \citet{Heckman:2016}, 
we obtain a scaling relation expressing the evolution of the outflow velocity by 
\begin{equation}
  \label{eq:po1-1}
  v\maxi \propto (1+z)^{2/3}.
\end{equation}
Equation (\ref{eq:po1-1}) is similar to Equation (\ref{eq:po1-0}), 
albait with a slight difference of the power.
This simple calculation suggests that 
the $v\maxi$ evolution originates from 
the $\Sigma_\SFR$ (i.e., size) evolution of galaxies.

\begin{figure}[!t]
  \epsscale{1.0}
  \plotone{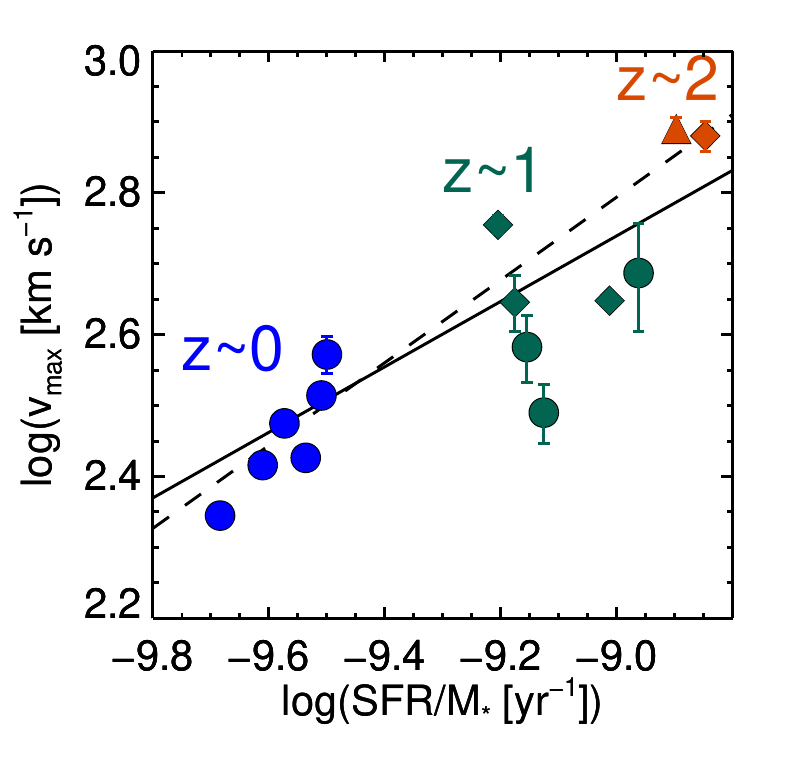}
  \caption{The outflow veocities as a function of $\SFR/M_*$.
    The top and bottom pannels show $v\maxi$ and $v\out$, respectively.
    The symbols and colors are the same as those in Figure \ref{fig:vwind}.
    In both panels, the green diamonds and the the orange triangles are 
    offset in $\SFR/M_*$ by 0.05 dex for clarity.
}
  \label{fig:ssfr}
\end{figure}

This interpretation also predicts 
a relation between $v\maxi$ and $\SFR/M_*$ at $z\sim0$--$2$.
At a fixed stellar mass of $\log(M_*/M_\odot) = 10.5$, 
\citet{Speagle:2014a} find 
\begin{equation}
\label{eq:po1-2}
\SFR/M_* \propto (1+z)^{2.8}.
\end{equation}
Using Equations (\ref{eq:po1-1})--(\ref{eq:po1-2}) and 
the relation of $v\maxi \propto \SFR^{1/3}$ at $z=0$ found by \citet{Heckman:2016}, 
we obtain the $v\maxi$ values increasing by 
\begin{equation}
  \label{eq:po1-25}
  v\maxi \propto (1+z)^{1.6}
\end{equation}
at a fixed stellar mass from $z\sim0$ to $2$.
Comparing Equations (\ref{eq:po1-2}) and (\ref{eq:po1-25}) 
finally yields
\begin{equation}
  \label{eq:po1-3}
  v\maxi \propto \SFR/M_*^{0.57}
\end{equation}
at a fixed stellar mass from $z\sim0$ to $2$.
Figure \ref{fig:ssfr} illustrates our fitting results of $v\maxi$ 
as a function of $\SFR/M_*$.
We fit all data points of the \zz-, \zo-, and \zt-samples 
with the form of $\log v\maxi \propto \beta\log\SFR/M_*$.
The best-fitting scaling factor is $\beta=0.46\pm0.02$ 
(solid line in Figure \ref{fig:ssfr}).
This $\beta$ value is simillar to the one of Equation (\ref{eq:po1-3}).
Moreover, if we use data points of only the \zz- and \zt-samples, 
we obtain $\beta = 0.58\pm0.02$ (dashed line in Figure \ref{fig:ssfr}) 
that is in excellent agreement with the one of Equation (\ref{eq:po1-3}). 
These results also support the interpretation that 
an increase in $v\maxi$ is caused by an increase in $\Sigma_\SFR$ of 
galaxies from $z\sim0$ to $2$.

\subsection{Mass Loading Factor}
\label{sec:mlf}
Another important parameter of the outflows is 
the mass loading factor $\eta$, 
which is defined by
\begin{equation}
  \label{eq:mlf1}
    \eta = \dot{M}\out/SFR, 
\end{equation}
where $\dot{M}\out$ is the mass outflow rate.
$\eta$ represents how the outflows contribute the feedback process of 
the galaxy formation and evolution.
Estimates of $\eta$ depend on assumptions of parameters.
Our aim is, under the set of the fiducial parameters, 
to compare our observational results with theoretical predictions 
on the redshift evolution of the mass loading factors 
\citep{Muratov:2015,Barai:2015,Mitra:2015}.

We estimate $\eta$ by following 
previous studies that adopt the ``down-the-barrel'' technique
\citep[e.g.,][]{Weiner:2009, Martin:2012, Rubin:2014}.
We use the absorption lines of \NaID, \MgI, and \CII
to calculate $\eta$ of the \zz-, \zo-, and \zt-samples, respectively.
Although IE (depth) of \CII is higher (deeper) than \NaID and \MgI, 
we directly compare the values of $\eta$
estimated from the lines. 
This is because Figure \ref{fig:vwind} shows that 
$v\out$ of \MgI and \MgII are comparable 
despite their different IE and depths of the absorption lines.

To estimate $\dot{M}\out$, 
we use the spherical flow model \citep{Rupke:2002, Martin:2012} that 
assumes the bi-conical outflow 
whose solid angle subtended by the outflowing gas is given by $\Omega$.
In the model, $\dot{M}\out$ is given by 
\begin{equation}
  \label{eq:mlf2}
  \dot{M}\out = \bar{m}_p \Omega C_f v\out R N(\eH),
\end{equation}
where $\bar{m}_p$ is mean atomic weight, 
$R$ is the inner radius of the outflows, 
and $N(\eH)$ is the column density of hydrogen.
We assume $\bar{m}_p=1.4$ amu and 
$\Omega=4\pi$ that is a case of a spherical outflow.
We also assume that $R$ is the same as the effective radius 
which are obtained from the 
$r_e$--$M_*$ relation of \citet{Shibuya:2015}.

We estimate $N(\eH)$ from the column density of an ion $\Xn$ that is expressed as
\begin{equation}
  \label{eq:mlf4}
  N(\Xn) = \frac{\tau_0b_\mathrm{D}}{1.497\times10^{-15}\lambda\sys f}, 
\end{equation}
where $f$ is the oscillator strength.
The oscillator strengths are taken from \citet{Morton:1991}.
We define the gas-phase abundance of an element X with respect to hydrogen 
as $(\eX/\eH)\sub{gas} \equiv N(\eX)/N(\eH)$, where
the column density of an element $N(\eX)$ is given by 
$N(X) = \sum_n N(\Xn)$.
$N(\Xn)$ can be converted into $N(\eH)$ with three parameters: 
the ionization fraction 
$\chi(\Xn) \equiv N(\Xn)/N(\eX)$, 
the dust depletion factor 
$d(\eX) \equiv (\eX/\eH)\sub{gas}/(\eX/\eH)\sub{c}$, 
and 
the cosmic metallicity at each redshift 
$\mu(\eX) \equiv (\eX/\eH)\sub{c}$, where the index c refers to the cosmic average.
We thus estimate $N(\eH)$ with 
\begin{equation}
  \label{eq:nh}
  N(\eH) = \frac{N(\Xn)}{\chi(\Xn)d(\eX)\mu(\eX)}.
\end{equation}

Because it is difficult to obtain these three parameters by observations, 
we adopt fiducial parameters for our calculations as our best estimate.
We use the solar metallicity for $\mu(\eX)$ of the \zz-sample, and 
a half of the solar metallicity for 
$\mu(\eX)$ of the \zo- and \zt-samples, 
i.e., $\mu(\eX) = 0.5 (\eX/\eH)$\msub{\odot}.
According to \citet{Morton:2003}, 
the solar metallicity values are 
$\log(\mathrm{Na/H})_\odot = -5.68$, 
$\log(\mathrm{Mg/H})_\odot = -4.42$, and 
$\log(\mathrm{C/H})_\odot = -3.48$.
The dust depletion factors are scaled to 
Milky Way values taken from \citet{Savage:1996}: 
$d(\mathrm{Na}) = 0.1$, 
$d(\mathrm{Mg}) = 0.03$, and 
$d(\mathrm{C})  = 0.4$.
For the ionization fractions, 
we take the moderate value of $\chi(\NaID)=0.1$, 
because photoionization models calculated by {\tt CLOUDY} 
\citep{Ferland:1998a, Ferland:2013a} suggest 
$\chi(\NaID) = 10^{-4}$--1 \citep{Murray:2007, Chisholm:2016a}.
Given a high ionization parameter at $z\gtrsim1$ \citep{Nakajima:2014},
we choose $\chi(\MgI)=0.05$, a half of $\chi(\NaID)$.
\citet{Chisholm:2016a} also find $\chi(\SiII) \simeq 0.2$.
Because the ionization energy of 
$\SiII$ (16 eV) and $\CII$ (24 eV) are comparable, 
we assume $\chi(\CII) = 0.2$ that is the same as $\SiII$.
In summary, the parameters for \NaID, \MgI, and \CII are 
($\chi$, $d$, $\mu$) = (0.1, 0.1, 2.1$\times 10^{-6}$), 
(0.1, 0.03, 1.9$\times 10^{-5}$), and 
(0.2, 0.4, 1.7$\times 10^{-4}$), respectively.
$N(\eH)$ is calculated by Equation (\ref{eq:nh}) 
with the parameter sets ($\chi$, $d$, $\mu$).

\begin{figure}[!t]
  \epsscale{1.0}
  \plotone{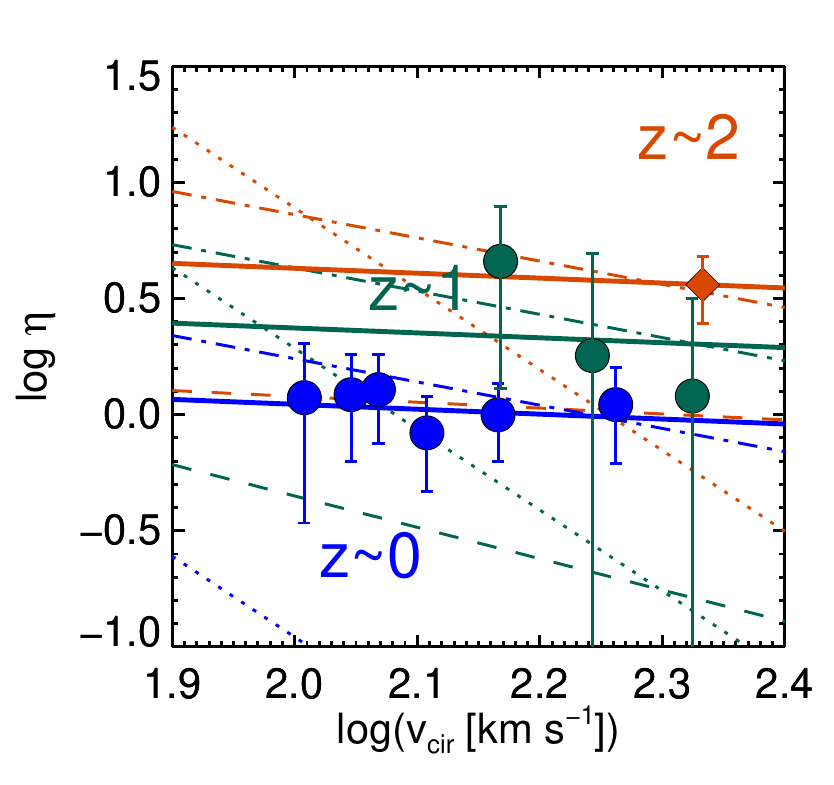}
  \caption{Mass loading factor as a function of $v_\mathrm{cir}$. 
  The data points are the same as those in Figure \ref{fig:vwind}.
  The colors of blue, green, and orange 
  correspond to $z\sim0$, 1, and 2, respectively.
  The solid lines denote the linear-square fitting results.
  The dot-dashed, dashed, and dotted lines indicate the models presented by 
  \citet{Muratov:2015}, \citet{Barai:2015}, and \citet{Mitra:2015}, respectively.}
  \label{fig:mlf}
\end{figure}

We estimate $\dot{M}\out$ by Equation (\ref{eq:mlf2}) 
with $C_f$ and $v\out$ that are derived in Section \ref{sec:ope}, 
and obtain $\eta$ by Equation (\ref{eq:mlf1}).
We find that our samples have $0 \lesssim \log\eta \lesssim 1$, 
which are consistent with the results of previous studies \citep[e.g.,][]{Heckman:2015}.
Figure \ref{fig:mlf} shows $\eta$ as a function of 
the halo circular velocity $v\cir$.
The values of $v\cir$ are calculated from $M_*$, 
with the $M_*$--$M_h$ relation in \citet{Behroozi:2013} and 
Equation (1) in \citet{Mo:2002}.
We perform the power-law fitting between $\eta$ and $v\cir$ of the \zz-sample, 
and obtain the best-fitting relation 
$\eta = \eta_1v\cir^a$ for 
$\eta_1 = \eEtaZz\pm\eEtaZzerr$ and $a=\eAZz\pm\eAZzerr$.
This scaling relation is consistent with the previous studies 
within the 1$\sigma$ uncertainties:  
$a=-0.98$ for strong outflows investigated 
with the UV observations of the local galaxies 
\citep{Heckman:2015} 
and 
$a=-1.0$ for $v_\mathrm{cir} > 60\ \mathrm{km\ s^{-1}}$ 
found with the FIRE simulations \citep{Muratov:2015}.
Our result, $a=\eAZz\pm\eAZzerr$, is weak evidence of a decreasing trend, 
although it is consistent with no correlation, i.e., $a=0$. 
Our result is consistent with the momentum-driven model ($a=-1$).
It is not conclusive, but we can rule out the energy-driven model ($a=-2$) 
at the 90 percentile significance level.
We also plot the data of the \zo- and \zt-samples in Figure \ref{fig:mlf}, 
but we cannot discuss the scaling relation of them due to few data points.

\subsection{Physical Origins of $\eta$ Evolution}
\label{sec:po}
Our observational results in Figure \ref{fig:mlf} indicate that 
$\eta$ increase from $z\sim0$ to 2 at a given circular velocity.
We fit the \zz-, \zo-, and \zt-samples with a power-law function $\eta = \eta_1v\cir^a$ 
at the fixed slope of $a=\eAZz$ 
that is the best-fit parameter of the \zz-sample (Section \ref{sec:mlf}).
We obtain the best-fit parameters 
$\eta_1 = \eEtaZz\pm \eEtaZzerrf$, $\eEtaZo\pm\eEtaZoerr$, and 
$\eEtaZt\pm\eEtaZterr$ 
for the \zz-, \zo-, and \zt-samples, 
which are illustrated in Figure \ref{fig:mlf} 
with the blue, green, and orange solid lines, respectively.
The redshift evolution of $\eta$ is obtained as $\eta \propto (1+z)^{\eZ\pm\eZerr}$
by a power-law fitting.

Theoretical methods predict an increase in the mass loading factors 
with increasing redshift \citep{Muratov:2015,Barai:2015,Mitra:2015}.
Our results reproduce the redshift evolution trend predicted 
by the theoretical studies.
In particular, the relation $\eta \propto (1+z)^{1.3}$ 
found by \citet{Muratov:2015} is 
in good agreement with our results (Figure \ref{fig:mlf}).
Since our estimates, however, include large uncertainties, 
we cannot constrain the theoretical models from the observations.

Some theoretical models suggest that 
a large amount of gas increases $v\out$, $v\maxi$, and $\eta$ 
towards high redshift. 
\citet{Barai:2015} claim that 
gas-rich galaxies at high redshift launch outflows with 
high $v\out$ and $\eta$.
Similarly, \citet{Hayward:2015} find that 
the value of $\eta$ increases exponentially 
by the increase {in} the gas fractions towards high redshift.

We discuss the redshift evolution of 
the outflows, SFR, and mass of cool gas in a galaxy.
If we assume that $\dot{M}\out$ is proportional to cool H{\sc i} 
gas mass $\cgas$, 
we can rewrite Equation (\ref{eq:mlf1}) as
\begin{equation}
  \label{eq:po1}
  \eta = \frac{\dot{M}\out}{\SFR} \propto \frac{\cgas v\out}{\SFR}.
\end{equation}
Hence the mass of the cool gas in the galaxy is 
\begin{equation}
  \label{eq:po2}
  \cgas \propto \frac{\eta \SFR}{v\out}.
\end{equation}
Below we calculate the redshift dependence of 
$\SFR, v\out, \eta$, and $\cgas$ 
at a fixed stellar mass $\log(M_*/M_\odot) = 10.5$.
The redshift evolution of $\SFR$ is given by Equation (\ref{eq:po1-2}).
If we assume that $v\out$ and $v\maxi$ follow the same dependence on $z$, 
the evolution of outflow velocities is expressed as Equation (\ref{eq:po1-25}).
We obtain 
$\eta \propto (1+z)^{\eZ\pm\eZerr}v\cir^{\eAZz\pm\eAZzerr}$, 
fitting the power-law functions 
to the results of Figure \ref{fig:mlf}.
Based on the estimates of 
the relation of $v\cir \propto (1+z)^{0.53\pm0.09}$ that 
we estimate from the $M_*$--$M_h$ relation in \citet{Behroozi:2013}, 
$\eta$ is written as $\eta \propto (1+z)^{1.1\pm0.7}$.
Substituting this equation, Equations (\ref{eq:po1-2}), and (\ref{eq:po1-25}) into 
Equation (\ref{eq:po2}), 
we obtain the redshift dependence of $\cgas$ by 
\begin{equation}
  \label{eq:po6}
  \cgas \propto (1+z)^{2.3\pm0.7}.
\end{equation}

Equation (\ref{eq:po6}) suggests that 
increasing $\cgas$ causes
the increases in 
outflow velocities, mass loading factors, and SFR 
with increasing redshift.
This increasing trend of $\cgas$ is 
consistent with independent observational results.
If we assume that the molecular gas mass is 
proportional to $\cgas$ at a given stellar mass, 
there is a relation of $\cgas \propto (1+z)^{2.7}$ 
obtained by the radio observations of \citet{Genzel:2015a}.
This relation is consistent with Equation (\ref{eq:po6}) 
within the 1$\sigma$ uncertainty.

As noted in Section \ref{sec:mlf}, however, 
the parameter sets used for deriving $\eta$ include large uncertainties.
The total uncertainty of $\chi(\Xn)$, $d(\eX)$, and $\mu(\eX)$ is 
$\gtrsim 0.5$ dex.
We thus think that 
the conclusion of an increase in $\eta$ is not strong.
However, we can securely claim that 
the theoretical models are consistent 
with our observational results 
under the assumption of the fiducial parameter sets shown 
in Section \ref{sec:mlf}.

\section{SUMMARY}
\label{sec:6}
We investigate redshift evolution of galactic outflows at $z\sim0$--2 
in the same stellar mass range.
We use rest-frame UV and optical spectra of star-forming galaxies at $z\sim0$, 1, and 2 
taken from the SDSS DR7, the DEEP2 DR4, and \citet{Erb:2006c, Erb:2006b}, respectively.
The outflows are identified with metal absorption lines: 
\NaID for the $z\sim0$ galaxies; 
\MgI and \MgII for the $z\sim2$ galaxies; and 
\CII and \CIV for the $z\sim2$ galaxies.
We construct composite spectra and 
measure the parameters of the galactic outflow properties 
such as the outflow velocities by the multi-component fitting 
with the aid of the SSP models.
Our results are summarized below.

\begin{enumerate}
\item We find that there are the scaling relations 
  between the outflow velocities and SFR at $z\sim0$: 
  for the maximum outflow velocity $v\maxi$ by 
  $v\maxi \propto \SFR^{\amZz\pm\amZzerr}$ and 
  for the central outflow velocity $v\out$ by 
  $v\out \propto \SFR^{\aoZz\pm\aoZzerr}$.
  The velocities $v\maxi$ and $v\out$ are consistent with previous studies.
\item We confirm that 
  both of the outflow velocities increase with increasing redshift.
  Because ions with higher ionization energy likely trace higher velocity clouds, 
  we compare ions with the similar ionization energies: 
  \NaID (IE $\simeq 5.1$ eV) and \MgI (IE $\simeq 7.6$ eV) from $z\sim0$ to 1, 
  and \MgII (IE $\simeq 15$ eV) and \CII (IE $\simeq 24$ eV) from $z\sim1$ to 2.
  The velocities $v\maxi$ and $v\out$ at $z\sim1$ ($z\sim2$) are 
  significantly higher than those at $z\sim0$ ($z\sim1$). 
\item The increase in the outflow velocities from $z\sim0$ to $2$ can 
    be explained by 
    the increase in $\Sigma_\mathrm{SFR}$ (i.e., a decrease of the galaxy size) 
    towards high redshift.
\item 
  To calculate the mass loading factors $\eta$, 
  we use \NaID at $z\sim0$, \MgI at $z\sim1$, and \CII at $z\sim2$.
  We then find that 
  the scaling relation between $\eta$ and the halo circular velocity $v\cir$ 
  is given by $\eta \propto v\cir^a$ for $a=\eAZz\pm\eAZzerr$ at $z\sim0$.
  The slope of $a=\eAZz\pm\eAZzerr$ suggests 
  that the outflows of the SDSS galaxies are launched by 
  a mechanism based on the momentum-driven model, which predicts $a=-1$.
\item We identify the increase in $\eta$ from $z\sim0$ to 2.
  The values of $\eta$ increase by $\eta \propto (1+z)^{\eZ\pm\eZerr}$, 
  with the fiducial parameter sets assumed in Section \ref{sec:mlf}.
  Note that the parameter sets include large uncertainties.
\item We find the increases in $v\out$, $v\maxi$, and $\eta$ towards high redshift by 
  observations. 
  These results are consistent with theoretical predictions that 
  explain the evolution by the increase in gas in high redshift galaxies.
\end{enumerate}

\acknowledgements

We thank the anonymous referee for constructive comments and suggestions.
We are grateful to C. N. A.  Willmer for sharing the measurements of 
the DEEP2 rest-frame magnitudes with us.

Funding for the SDSS and SDSS-II has been provided by the Alfred P. Sloan Foundation, the Participating Institutions, the National Science Foundation, the U.S. Department of Energy, the National Aeronautics and Space Administration, the Japanese Monbukagakusho, the Max Planck Society, and the Higher Education Funding Council for England. The SDSS Web Site is \url{http://www.sdss.org/}.
The SDSS is managed by the Astrophysical Research Consortium for the Participating Institutions. The Participating Institutions are the American Museum of Natural History, Astrophysical Institute Potsdam, University of Basel, University of Cambridge, Case Western Reserve University, University of Chicago, Drexel University, Fermilab, the Institute for Advanced Study, the Japan Participation Group, Johns Hopkins University, the Joint Institute for Nuclear Astrophysics, the Kavli Institute for Particle Astrophysics and Cosmology, the Korean Scientist Group, the Chinese Academy of Sciences (LAMOST), Los Alamos National Laboratory, the Max-Planck-Institute for Astronomy (MPIA), the Max-Planck-Institute for Astrophysics (MPA), New Mexico State University, Ohio State University, University of Pittsburgh, University of Portsmouth, Princeton University, the United States Naval Observatory, and the University of Washington.

Some of the data presented herein were obtained at the W. M. Keck Observatory, 
which is operated as a scientific partnership 
among the California Institute of Technology, 
the University of California and the National Aeronautics and Space Administration. 
The Observatory was made possible by the generous financial support 
of the W. M. Keck Foundation. 
The DEEP2 team and Keck Observatory acknowledge the very significant cultural role 
and reverence that the summit of Mauna Kea has always had within the indigenous Hawaiian 
community and appreciate the opportunity to conduct observations from this mountain.
The analysis pipeline used to reduce the DEIMOS data was developed 
at UC Berkeley with support from NSF grant AST-0071048. 
Funding for the DEEP2 Galaxy Redshift Survey has been
provided by NSF grants AST-95-09298, AST-0071048, AST-0507428, and AST-0507483 
as well as NASA LTSA grant NNG04GC89G.
This research has made use of the Keck Observatory Archive (KOA), 
which is operated by the W. M. Keck Observatory and 
the NASA Exoplanet Science Institute (NExScI), 
under contract with the National Aeronautics and Space Administration.

This work is supported by 
World Premier International Research Center Initiative (WPI Initiative), MEXT, Japan, 
and KAKENHI (15H02064) Grant-in-Aid for Scientific Research (A) through Japan Society 
for the Promotion of Science (JSPS).

\bibliographystyle{aasjournal}
\bibliography{$HOME/Documents/set_TeX/reference} 

\end{document}

%% file: tb1.tex
\tabletypesize{\scriptsize} 
\begin{deluxetable*}{llccccccc}
\tablecolumns{9}
\tablewidth{0pt}
\tablecaption{Properties of the stacked spectra \label{tb:1}}
\centering
\tablehead{
\colhead{sample} & \colhead{line} & \colhead{number} & \colhead{$z$} & \colhead{SFR} & \colhead{${M_*}$} & \colhead{${v\out}$} & \colhead{${v\maxi}$} & \colhead{${\eta}$} \\
\colhead{} & \colhead{} & \colhead{} & \colhead{} & \colhead{($\mathrm{M_\odot\ yr^{-1}}$)} & \colhead{($\mathrm{M_\odot}$)} & \colhead{($\mathrm{km\ s^{-1}}$)} & \colhead{($\mathrm{km\ s^{-1}}$)} & \colhead{}
}
\startdata
\zz-sample 
 & \NaID & 126 & 0.064 &  0.49 & 10.2 & 146 $\pm$ 5.2 & 221 $\pm$ 18 &  1.2 $\pm$ 0.84 \\ 
 & \NaID & 113 & 0.075 &  0.69 & 10.3 & 152 $\pm$ 4.4 & 261 $\pm$ 12 &  1.2 $\pm$ 0.59 \\ 
 & \NaID & 109 & 0.085 &  0.80 & 10.4 & 160 $\pm$ 4.5 & 299 $\pm$ 11 &  1.3 $\pm$ 0.53 \\ 
 & \NaID & 138 &  0.11 &  0.93 & 10.5 & 144 $\pm$ 4.3 & 267 $\pm$ 11 & 0.83 $\pm$ 0.37 \\ 
 & \NaID & 141 &  0.13 &   1.1 & 10.6 & 153 $\pm$ 4.3 & 327 $\pm$ 10 &  1.0 $\pm$ 0.37 \\ 
 & \NaID & 123 &  0.14 &   1.3 & 10.8 & 165 $\pm$ 9.7 & 373 $\pm$ 20 &  1.1 $\pm$ 0.49 \\ 
\zo-sample
 &  \MgI  & 662 & 1.4 & 1.0 & 9.99 & 220 $\pm$  38 & 486 $\pm$ 64 &  4.6 $\pm$  3.3 \\
 &  \MgI  & 394 & 1.4 & 1.3 & 10.4 & 164 $\pm$  16 & 309 $\pm$ 91 &  1.8 $\pm$  3.2 \\
 &  \MgI  & 277 & 1.4 & 1.5 & 10.6 & 175 $\pm$  19 & 382 $\pm$ 98 &  1.2 $\pm$  2.0 \\
 &  \MgII & 662 & 1.4 & 1.0 & 9.99 & 207 $\pm$ 5.0 & 445 $\pm$  9 & \ldots \\
 &  \MgII & 394 & 1.4 & 1.3 & 10.4 & 180 $\pm$  16 & 442 $\pm$ 27 & \ldots \\
 &  \MgII & 277 & 1.4 & 1.5 & 10.6 & 241 $\pm$ 7.1 & 569 $\pm$ 12 & \ldots \\
\zt-sample
 &  \CII & 25 & 2.2 & 1.4 & 10.3 & 428 $\pm$ 20 & 759$\pm$ 20 & 3.6 $\pm$ 1.2 \\
 &  \CIV & 25 & 2.2 & 1.4 & 10.3 & 445 $\pm$ 16 & 776$\pm$ 16 & \ldots \\
\enddata
\tablenotetext{a}{Number of galaxies used for a composite spectrum.}
\tablenotetext{b}{Median redshift.}
\tablenotetext{c}{Median global (i.e., aperture corrected) star formation rate. }
\tablenotetext{d}{Median stellar mass. }
\tablenotetext{e}{Central outflow velocity defined in Section \ref{sec:outflowvelocity}.}
\tablenotetext{f}{Maximum outflow velocity defined in Section \ref{sec:outflowvelocity}.}
\tablenotetext{g}{Mass loading factor defined in Section \ref{sec:mlf}.}

\end{deluxetable*}